%% file: promptDs.tex
\documentclass[ALICE,manyauthors]{cernphprep}
\usepackage[comma,square,numbers,sort&compress]{natbib}
\usepackage[T1]{fontenc}
\usepackage{hyperref}
\usepackage{lineno}
\usepackage{xcolor}
\usepackage{xspace}
\usepackage{multirow}
\usepackage{makecell}
\usepackage{booktabs}
\usepackage{upgreek}
\begin{document}
\input{commands.tex}

\begin{titlepage}
\PHyear{2021}       
\PHnumber{187}      
\PHdate{16 September}  

\title{Measurement of prompt $\pmb{\Ds}$-meson production and azimuthal anisotropy in Pb--Pb collisions at $\pmb{\sqrtsNN = 5.02~\TeV}$}
\ShortTitle{Prompt $\Ds$ mesons in Pb--Pb at $\sqrtsNN = 5.02~\TeV$}   

\Collaboration{ALICE Collaboration\thanks{See Appendix~\ref{app:collab} for the list of collaboration members}}
\ShortAuthor{ALICE Collaboration} 

\begin{abstract}
The production yield and angular anisotropy of prompt $\Ds$ mesons were measured as a function of transverse momentum ($\pt$) in Pb--Pb collisions at a centre-of-mass energy per nucleon pair $\sqrtsNN=5.02~\tev$ collected with the ALICE detector at the LHC.
$\Ds$ mesons and their charge conjugates were reconstructed at midrapidity ($|y|<0.5$) from their hadronic decay channel $\Dstophipi$, with $\phitoKK$,  
in the $\pt$ intervals $2<\pt<50~\GeV/c$ and $2<\pt<36~\GeV/c$ for the 0--10\% and 30--50\% centrality intervals. 
For $\pt>10~\gev/c$, the measured $\Ds$-meson nuclear modification factor $\RAA$ is consistent with the one of non-strange D mesons within uncertainties, while at lower $\pt$ a hint for a $\Ds$-meson $\RAA$ larger than that of non-strange D mesons is seen. 
The enhanced production of $\Ds$ relative to non-strange D mesons is also studied by comparing the $\pt$-dependent $\Ds/\Dzero$ production yield ratios in Pb--Pb and in pp collisions. The ratio measured in Pb--Pb collisions is found to be on average higher than that in pp collisions in the interval $2<\pt<8~\GeV/c$ with a significance of $2.3\sigma$ and $2.4\sigma$ for the 0--10\% and 30--50\% centrality intervals.
The azimuthal anisotropy coefficient $v_2$ of prompt $\Ds$ mesons was measured in Pb--Pb collisions in the 30--50\% centrality interval and is found to be compatible with that of non-strange D mesons.
The main features of the measured $\RAA$, $\Ds/\Dzero$ ratio, and $v_2$ as a function of $\pt$ are described by theoretical calculations of charm-quark transport in a hydrodynamically expanding quark--gluon plasma including hadronisation via charm-quark recombination with light quarks from the medium. The $\pt$-integrated production yield of $\Ds$ mesons is compatible with the prediction of the statistical hadronisation model.

\end{abstract}
\end{titlepage}

\setcounter{page}{2} 

\input{intro}
\input{detectoranddatasample}
\input{analysis}
\input{systematics}
\input{results}
\input{conclusions}
%

\newenvironment{acknowledgement}{\relax}{\relax}
\begin{acknowledgement}
\section*{Acknowledgements}
\input{fa_2021-08-20.tex}
\end{acknowledgement}

\bibliographystyle{utphys}   
\bibliography{bibliography}

\newpage
\appendix

%
%

\section{The ALICE Collaboration}
\label{app:collab}
\input{2021-08-20-Alice_Authorlist_2021-08-20.tex}  
\end{document}

%% file: commands.tex
%

\newcommand{\ee}               {\mathrm{e^+e^-}} 
\newcommand{\pp}               {\mathrm{pp}}
\newcommand{\ppbar}            {\mathrm{p\overline{p}}}
\newcommand{\pPb}              {\mathrm{p--Pb}}
\newcommand{\PbPb}             {\mathrm{Pb--Pb}}

\newcommand{\av}[1]            {\left\langle #1 \right\rangle}

\newcommand{\s}                {\sqrt{s}}
\newcommand{\sqrtsNN}          {\sqrt{s_\mathrm{NN}}}
\newcommand{\Npart}            {N_\mathrm{part}}
\newcommand{\Ncoll}            {N_\mathrm{coll}}
\newcommand{\RpPb}             {R_\mathrm{pPb}}
\newcommand{\RAA}              {R_\mathrm{AA}}
\newcommand{\TAA}              {T_\mathrm{AA}}
\newcommand{\pt}               {p_\mathrm{T}}
\newcommand{\mt}               {m_\mathrm{T}}
\newcommand{\de}               {\mathrm{d}}
\newcommand{\dEdx}             {\de E/\de x}
\newcommand{\dNdpt}            {\de N/\de\pt}
\newcommand{\dNdy}             {\de N/\de y}
\newcommand{\mur}              {\mu_\mathrm{R}}
\newcommand{\muf}              {\mu_\mathrm{F}}
\newcommand{\T}                {\mathrm{T}}
\newcommand{\meanpt}           {\langle p_\mathrm{T} \rangle}
\newcommand{\sigmatot}         {\sigma_{\rm tot}}
\newcommand{\fprompt}          {f_\mathrm{prompt}}
\newcommand{\fnonprompt}       {f_\mathrm{non\text{-}prompt}}
\newcommand{\f}[1]             {f_\mathrm{#1}}
\newcommand{\rawY}[1]          {Y_{#1}}
\newcommand{\effNP}            {(\mathrm{Acc}\times\epsilon)_{\mathrm{non\text{-}prompt}}}
\newcommand{\effP}             {(\mathrm{Acc}\times\epsilon)_{\mathrm{prompt}}}
\newcommand{\Np}               {N_\mathrm{prompt}}
\newcommand{\Nnp}              {N_\mathrm{non\text{-}prompt}}
\newcommand{\qTPC}{q_2^{\rm TPC}}
\newcommand{\qVZEROA}{q_2^{\rm V0A}}
\newcommand{\vtwo}{v_\mathrm{2}}
\newcommand{\vtwotot}{v^\mathrm{tot}_\mathrm{2}}
\newcommand{\vtwosig}{v^\mathrm{sig}_\mathrm{2}}
\newcommand{\vtwobkg}{v^\mathrm{bkg}_\mathrm{2}}
\newcommand{\vtwodplus}{v^\mathrm{D^{+}}_\mathrm{2}}
\newcommand{\vtwoprompt}{v^\mathrm{prompt}_\mathrm{2}}
\newcommand{\vtwofd}{v^\mathrm{non\text{-}prompt}_\mathrm{2}}
\newcommand{\Nsig}{N^\mathrm{sig}}
\newcommand{\Nbkg}{N^\mathrm{bkg}}
\newcommand{\Ndplus}{N^\mathrm{D^{+}}}
\newcommand{\Qtwo}{Q_\mathrm{2}}
\newcommand{\Rtwoflow}{R_{\rm 2}}

\newcommand{\eV}               {\mathrm{eV}}
\newcommand{\keV}              {\mathrm{keV}}
\newcommand{\MeV}              {\mathrm{MeV}}
\newcommand{\GeV}              {\mathrm{GeV}}
\newcommand{\TeV}              {\mathrm{TeV}}
\newcommand{\ev}               {\mathrm{eV}}
\newcommand{\kev}              {\mathrm{keV}}
\newcommand{\mev}              {\mathrm{MeV}}
\newcommand{\mevc}             {\mathrm{MeV}/c}
\newcommand{\mevcsquared}      {\mathrm{MeV}/c^2}
\newcommand{\gev}              {\mathrm{GeV}}
\newcommand{\gevc}             {\mathrm{GeV}/c}
\newcommand{\tev}              {\mathrm{TeV}}
\newcommand{\fm}               {\mathrm{fm}}
\newcommand{\mm}               {\mathrm{mm}} 
\newcommand{\cm}               {\mathrm{cm}}
\newcommand{\m}                {\mathrm{m}}
\newcommand{\mum}              {\mathrm{\upmu m}}
\newcommand{\ns}               {\mathrm{ns}}
\newcommand{\mrad}             {\mathrm{mrad}}
\newcommand{\mb}               {\mathrm{mb}}
\newcommand{\mub}              {\mathrm{\upmu b}}
\newcommand{\lumi}             {\mathcal{L}_\mathrm{int}}
\newcommand{\nbinv}            {\mathrm{nb^{-1}}}

\newcommand{\ITS}              {\mathrm{ITS}}
\newcommand{\TOF}              {\mathrm{TOF}}
\newcommand{\ZDC}              {\mathrm{ZDC}}
\newcommand{\ZDCs}             {\mathrm{ZDC}}
\newcommand{\ZNA}              {\mathrm{ZNA}}
\newcommand{\ZNC}              {\mathrm{ZNC}}
\newcommand{\SPD}              {\mathrm{SPD}}
\newcommand{\SDD}              {\mathrm{SDD}}
\newcommand{\SSD}              {\mathrm{SSD}}
\newcommand{\TPC}              {\mathrm{TPC}}
\newcommand{\TRD}              {\mathrm{TRD}}
\newcommand{\VZERO}            {\mathrm{V0}}
\newcommand{\VZEROA}           {\mathrm{V0A}}
\newcommand{\VZEROC}           {\mathrm{V0C}}

\newcommand{\pip}              {\mathrm{\uppi^{+}}}
\newcommand{\pim}              {\mathrm{\uppi^{-}}}
\newcommand{\kap}              {\mathrm{\rm{K}^{+}}}
\newcommand{\kam}              {\mathrm{\rm{K}^{-}}}
\newcommand{\pbar}             {\mathrm{\rm\overline{p}}}
\newcommand{\kzero}            {\mathrm{K^0_S}}
\newcommand{\lmb}              {\mathrm{\Lambda}}
\newcommand{\almb}             {\mathrm{\overline{\Lambda}}}
\newcommand{\Om}               {\mathrm{\Omega^-}}
\newcommand{\Mo}               {\mathrm{\overline{\Omega}^+}}
\newcommand{\X}                {\mathrm{\Xi^-}}
\newcommand{\Ix}               {\mathrm{\overline{\Xi}^+}}
\newcommand{\Xis}              {\mathrm{\Xi^{\pm}}}
\newcommand{\Oms}              {\mathrm{\Omega^{\pm}}}
\newcommand{\DzerotoKpi}       {\mathrm{D^0 \to K^-\uppi^+}}
\newcommand{\DplustoKpipi}     {\mathrm{D^+\to K^-\uppi^+\uppi^+}}
\newcommand{\DstartoDpi}       {\mathrm{D^{*+} \to \rm D^0 \uppi^+}}
\newcommand{\Dstophipi}        {\mathrm{D_s^+\to \upphi\uppi^+}}
\newcommand{\Dstophipipm}      {\mathrm{D_s^\pm\to \upphi\uppi^\pm}}
\newcommand{\DstophipitoKKpi}  {\mathrm{D_s^+\to \upphi\uppi^+\to K^-K^+\uppi^+}}
\newcommand{\DplustoKKpi}      {\mathrm{D^+\to K^-K^+\uppi^+}}
\newcommand{\phitoKK}          {\mathrm{\upphi\to  K^-K^+}}
\newcommand{\DstoKzerostarK}   {\mathrm{D_s^+\to \overline{K}^{*0} K^+}}
\newcommand{\Dstofzeropi}      {\mathrm{D_s^+\to f_0(980) \uppi^+}}
\newcommand{\fzero}            {\mathrm{f_0(980)}}
\newcommand{\Kzerostar}        {\mathrm{\overline{K}^{*0}}}
\newcommand{\Dzero}            {\mathrm{D^0}}
\newcommand{\Dzerobar}         {\mathrm{\overline{D}\,^0}}
\newcommand{\Dstar}            {\mathrm{D^{*+}}}
\newcommand{\Dstarm}           {\mathrm{D^{*-}}}
\newcommand{\DstarZero}        {\mathrm{D^{*0}}}
\newcommand{\DstarS}           {\mathrm{D_s^{*+}}}
\newcommand{\Dplus}            {\mathrm{D^+}}
\newcommand{\Dminus}           {\mathrm{D^-}}
\newcommand{\Ds}               {\mathrm{D_s^+}}
\newcommand{\Dspm}             {\mathrm{D_s^\pm}}
\newcommand{\Dsstar}           {\mathrm{D_s^{*+}}}
\newcommand{\KKpi}             {\mathrm{K^-K^+\uppi^+}}
\newcommand{\cubar}            {\mathrm{c\bar{u}}}
\newcommand{\cdbar}            {\mathrm{c\bar{d}}}
\newcommand{\ccbar}            {\mathrm{c\overline{c}}}
\newcommand{\bbbar}            {\mathrm{b\overline{b}}}
\newcommand{\Bzero}            {\mathrm{B^0}}
\newcommand{\Bplus}            {\mathrm{B^+}}
\newcommand{\Bzeroplus}        {\mathrm{B^{0,+}}}
\newcommand{\Bs}               {\mathrm{B_s^0}}
\newcommand{\Lambdab}          {\mathrm{\Lambda_b^0}}
\newcommand{\Jpsi}             {\mathrm{J}/\uppsi}
\newcommand{\Vdecay} 	       {\mathrm{V^{0}}}
\newcommand{\bhad}             {\mathrm{H_b}}
\newcommand{\Ztobbbar}         {\mathrm{Z\to b\overline{b}}}
\newcommand{\fctoD}            {f(\mathrm{c}\to\mathrm{D})}
\newcommand{\fbtoB}            {f(\mathrm{b}\to\mathrm{B})}
\newcommand{\fctoHc}           {f(\mathrm{c}\to\mathrm{H_c})}
\newcommand{\fbtoHb}           {f(\mathrm{b}\to\mathrm{H_b})}

%% file: intro.tex
\section{Introduction}
\label{sec:intro}

Strongly-interacting matter at temperatures exceeding the pseudo-critical value of $T_{\rm pc} \approx 154\text{--}158~\mev$ and at vanishing baryon density is predicted to behave as a plasma of deconfined quarks and gluons (QGP)~\cite{Bazavov:2018mes,Borsanyi:2020fev}.
A QGP is formed and studied in ultrarelativistic heavy-ion collisions at the CERN Large Hadron Collider (LHC) and existing measurements indicate that it behaves as a strongly-coupled liquid-like system~\cite{Busza:2018rrf}.
The lifetime of the QGP produced at the energy densities reached at the LHC is of the order of 10 $\mathrm{fm}/c$~\cite{Aamodt:2011mr}.
Heavy quarks (charm and beauty) are sensitive probes to investigate the
properties of the medium formed in these collisions.
Due to their large masses, heavy quarks are produced predominantly
in hard partonic scattering processes occurring during the early stages of the
collision (i.e.\ on timescales shorter than the QGP formation time) and therefore experience the entire evolution of the medium.
Heavy quarks propagate through the expanding hot and dense medium, interacting and exchanging energy and momentum with its constituents via both inelastic and elastic quantum chromodynamic (QCD) processes.
At high momentum, the main effect of these interactions is the energy loss of the heavy quarks in the QGP due to medium-induced gluon radiation and collisional processes.
On the other hand, low-momentum heavy quarks, including those shifted to low momentum by the energy loss, probe the diffusion regime dominated by elastic interactions.
Since the charm and beauty quark masses are large compared to the medium temperature, the propagation of low-momentum heavy quarks through the fireball can be treated as a ``Brownian motion'', characterised by many elastic collisions with relatively small momentum transfers~\cite{Moore:2004tg,Rapp:2008qc}.
As a consequence of the large number of soft collisions with the medium constituents, heavy quarks can acquire significant collective flow when diffusing through the expanding fireball.
Due to their large masses, charm quarks have a thermalisation time which is comparable to the fireball lifetime~\cite{Batsouli:2002qf,Moore:2004tg}, and therefore they carry sensitive information on their coupling strength to the expanding medium, preserving memory of the thermalisation process.
The process of hadronisation is also predicted to be modified in the presence of the QGP. Once the fireball approaches the pseudo-critical temperature for the transition to a hadron gas, a significant fraction of low- and intermediate-momentum heavy quarks could hadronise via recombination with other quarks from the medium~\cite{Fries:2003vb,Greco:2003xt,Greco:2003vf,Ravagli:2007xx}, in competition with the fragmentation mechanism, which describes quark-to-hadron transitions in pp, $\mathrm{e}^\pm$p, and $\mathrm{e}^+\mathrm{e}^-$ collisions~\cite{Lisovyi:2015uqa,Acharya:2021set}.

The effects of the interaction of heavy quarks with the medium are commonly quantified by two main observables: the nuclear modification factor $\RAA$ and the elliptic flow $v_2$. The $\RAA$ is defined as the ratio of the transverse-momentum ($\pt$) differential yields in nucleus--nucleus (AA) collisions and the cross section in proton--proton collisions, scaled by the average nuclear overlap function 
$\av{T_{\rm AA}}$
\begin{equation}
\label{eq:Raa}
R_{\rm AA}(\pt)=
{\frac{1}{\av{T_{\rm AA}}}} \times 
{\frac{{\rm d} N_{\rm AA}/{\rm d}\pt}{{\rm d}\sigma_{\rm pp}/{\rm d}\pt}}\,,
\end{equation}
where the yield in nucleus--nucleus collisions ${\rm d} N_{\rm AA}/{\rm d}\pt$ is measured in a given centrality interval and the $\av{T_{\rm AA}}$ value is proportional to the average number of nucleon--nucleon collisions~\cite{Wang:1998bha}. The $\av{T_{\rm AA}}$ can be estimated via Glauber-model calculations tuned to match the measured multiplicity distribution of charged particles~\cite{Miller:2007ri}.
The elliptic flow $v_2$ is the second coefficient of the Fourier expansion of the particle-yield distribution in the azimuthal direction $\varphi$ relative to the initial-state symmetry plane angle $\Psi_2$:  $v_2 = \langle {\rm cos}[2(\varphi-\Psi_2)]\rangle$, where $\langle \rangle$ indicates the average over all particles and all events~\cite{Ollitrault:1992bk,Poskanzer:1998yz}.

Measurements of non-strange D-meson production in heavy-ion collisions at RHIC~\cite{Adam:2018inb} and LHC~\cite{Sirunyan:2017xss,Acharya:2018hre,RaaD} energies show a substantial suppression of the D-meson yields compared to pp collisions at intermediate and high $\pt$.
In central nucleus--nucleus collisions, the $\RAA$ exhibits a pronounced drop for $\pt>4\text{--}5~\GeV/c$, reaches a minimum around $\pt \approx 8~\GeV/c$, and slightly increases at higher $\pt$.
This trend is described by different state-of-the-art model calculations of charm-quark energy loss in the QGP~\cite{Xu:2014tda,Stojku:2020tuk,Kang:2016ofv}.
A positive D-meson $v_2$ is measured at $\pt > 8\text{--}10~\GeV/c$ for semicentral Pb--Pb collisions at the LHC~\cite{Sirunyan:2017plt,Acharya:2020pnh}, and it is understood as originating from the path-length dependence of the charm-quark energy loss in the geometrically anisotropic medium created in collisions with finite impact parameter.
At lower $\pt$, larger values of D-meson $v_2$ are observed accompanied by a bump-like structure in the $\RAA$ reflecting the radial flow of the fireball~\cite{Acharya:2018hre,Sirunyan:2017xss,RaaD}.
In particular, the D-meson $v_2$ for semicentral collisions shows a maximum value at $\pt \approx 3~\gev/c$, a clear mass ordering $v_2({\rm D})<v_2({\rm p})<v_2({\pi})$ at low $\pt$ ($\pt < 3~\gev/c$), and  a similar magnitude as the $v_2$ of charged pions at intermediate $\pt$ ($3<\pt<6~\gev/c$)~\cite{Acharya:2020pnh,Sirunyan:2017plt}.
These features are consistent with a scenario in which low-momentum charm quarks acquire a significant collective flow when diffusing through the expanding QGP and hadronise via recombination with light quarks from the medium.\ 
The measured $\RAA$ and $v_2$ in this $\pt$ region are described qualitatively, and to some extent also quantitatively, by transport models including charm-quark interactions in a hydrodynamically expanding QGP and hadronisation via both fragmentation and recombination~\cite{Nahrgang:2013xaa,Katz:2019fkc,Beraudo:2014boa,Beraudo:2017gxw,Cao:2016gvr,Cao:2017hhk,He:2019vgs,Li:2019lex,Scardina:2017ipo,Plumari:2019hzp,Ke:2018jem,Song:2015sfa}. However, a simultaneous description of the nuclear modification factor and the anisotropic flow of D mesons is still a challenge for theoretical models.

Studies of the production of different charm-hadron species, dubbed heavy-flavour hadrochemistry, can provide information on the hadronisation mechanism of charm quarks.
In particular, an enhancement of the ground-state charm-strange meson yield relative to that of non-strange D mesons is expected in nucleus--nucleus collisions  at low and intermediate momenta as compared to pp interactions, if the dominant process for D-meson  formation is the recombination of charm quarks with light quarks from the medium, due to the large abundance of strange quarks in the QGP~\cite{Rafelski:1982pu,Koch:1986ud,Andronic:2003zv,Kuznetsova:2006bh,He:2012df}.
It was also pointed out in Ref.~\cite{He:2012df} that the comparison of the $v_2$ of $\Ds$ mesons to that of D mesons without strange-quark content ($\Dzero$, $\Dplus$, and $\Dstar$) could provide sensitivity to the transport properties of the hadronic phase, since $\Ds$ mesons are expected to decouple early from the hadron gas and therefore do not pick up significant additional $v_2$ in the hadronic phase.

The production of $\Ds$ mesons was measured at RHIC~\cite{Adam:2021qty} and the LHC~\cite{Adam:2015jda,Acharya:2018hre} in Au--Au and Pb--Pb collisions at different centralities. 
So far, the results have shown that at low and intermediate $\pt$ the $\Ds/\Dzero$ ratio in central, semicentral, and peripheral collisions is larger than the value measured in pp collisions, though the relatively large uncertainties do not allow firm conclusions.
The magnitude and the $\pt$ dependence of the $\Ds/\Dzero$ ratio are captured, at least qualitatively, by models including hadronisation via quark coalescence along with strangeness enhancement in the QGP~\cite{Zhao:2018jlw,Plumari:2017ntm,He:2012df,He:2019vgs}, suggesting a relevant role of recombination processes in the hadronisation of low-momentum charm quarks in the QGP.

In this Letter, we report the measurements of the $\pt$-differential yield and the nuclear modification factor of prompt $\Ds$ mesons in central (0--10$\%$) and semicentral (30--50$\%$) Pb--Pb collisions at $\sqrtsNN = 5.02~\TeV$, together with the measurement of the prompt $\Ds$-meson elliptic flow in semicentral collisions.
$\Ds$ mesons and their charge conjugates were reconstructed at midrapidity, $|y|<0.5$, through their hadronic decay channel $\Dstophipi$ with a subsequent decay $\phitoKK$.
Prompt $\Ds$ mesons are defined as those produced directly in the hadronisation of charm quarks or originating from the decays of directly-produced excited open-charm and charmonium states, hence excluding weak decays of beauty hadrons.
The data sample used for the analysis reported in this paper was collected with the ALICE detector at the end of 2018 and is larger by a factor of about 8 (4) for central (semicentral) collisions with respect to the sample collected in 2015, used for the previous publications of $\Ds$-meson $\RAA$ and $v_2$~\cite{Acharya:2018hre,Acharya:2017qps}.

%% file: detectoranddatasample.tex
\section{Experimental apparatus and data sample}
\label{sec:detector}

The ALICE apparatus comprises a central barrel, which is composed of a set of detectors for charged particle reconstruction and identification at midrapidity, a forward muon spectrometer, and various forward and backward
detectors for triggering and event characterisation.
A detailed description of the detectors and an overview of their typical performances can be found in Refs.~\cite{Aamodt:2008zz,Abelev:2014ffa}.

The $\Ds$-meson decay candidates and charged conjugates were reconstructed and identified with the central barrel detectors, which cover the full azimuth in the pseudorapidity interval $|\eta| < 0.9$ and are embedded in a large solenoidal magnet providing a homogeneous magnetic field $B = 0.5~\mathrm{T}$ parallel to the beam direction.
Charged-particle trajectories are reconstructed from their hits in the Inner Tracking System (ITS) and the Time Projection Chamber (TPC).
The ITS is the innermost detector of the ALICE central barrel, it consists of six cylindrical layers of silicon detectors, allowing a precise determination of the track parameters in the vicinity of the interaction point.
The TPC provides track reconstruction with up to 159 three-dimensional space points along the trajectory of a charged particle and provides particle identification via the measurement of the specific ionisation energy loss $\dEdx$.
The Time-Of-Flight (TOF) detector, positioned at a radial distance of about 4 m from the beam axis, extends the particle-identification capabilities of the TPC by measuring the flight time of the charged particles from the interaction point to the TOF.
The V0 detector is used for triggering and event selection, as well as for the estimation of the collision centrality and the reference plane for the elliptic flow measurement.
It consists of two scintillator arrays, located on both sides of the nominal interaction point and covering the full azimuth in the pseudorapidity intervals $-3.7< \eta <-1.7$ (V0C) and $2.8< \eta <5.1$ (V0A).
The neutron Zero Degree Calorimeters (ZDC), located along the beam axis on both sides of the central barrel at about 110 m distance from the interaction point, are used for event selection, along with the V0 detector.

The events used in the analysis were recorded with a minimum bias (MB) trigger which required coincident signals in the V0A and V0C detectors. 
Two additional trigger classes were used to enrich the sample of central and semicentral collisions via an online event selection based on the V0-signal amplitude.
Background events due to the interaction of one of the beams with residual gas in the vacuum tube and other machine-induced backgrounds were rejected offline using the V0 and the ZDC timing information~\cite{Abelev:2014ffa}.
In order to have a uniform acceptance in pseudorapidity, only events with a primary vertex reconstructed within $\pm10$~cm from the centre of the detector along the beam-line direction were considered in the analysis.
Collisions were classified into centrality intervals, defined in terms of percentiles of the hadronic Pb--Pb cross section, based on the V0 signal amplitude, as described in detail in Ref.~\cite{Adam:2015ptt}.
Central and semicentral collisions were considered in the analysis of the $\Ds$-meson production.
The sample of central collisions consists of about $100 \times 10^6$ events in the 0--10\% centrality interval, corresponding to an integrated luminosity $\mathcal{L}_{\rm int}~\simeq 130~\upmu \mathrm{b}^{-1}$.
For semicentral collisions, a sample of about $85 \times 10^6$ events in the 30--50\% interval was utilised, corresponding to $\mathcal{L}_{\rm int}~\simeq 56~\upmu \mathrm{b}^{-1}$.
The average values of the nuclear overlap function, $\av{\TAA}$, for the considered central and semicentral event intervals were estimated via Glauber-model simulations anchored to the measured charged-particle multiplicity distribution, and are $23.26\pm0.17$ mb$^{-1}$ and $3.92\pm0.06$ mb$^{-1}$~\cite{ALICE-PUBLIC-2018-011}, respectively.

The Monte Carlo samples utilised in the analysis were obtained simulating Pb--Pb collisions with the HIJING~1.36~\cite{PhysRevD.44.3501} event generator. In each simulated event, additional $\ccbar$- and $\bbbar$-quark pairs were injected using the PYTHIA~8.243 event generator~\cite{Sjostrand:2006za, Sjostrand:2014zea} (Monash-13 tune~\cite{Skands:2014pea}) and $\Ds{}$ mesons were forced to decay into the hadronic channel of interest for the analysis. The generated particles were propagated through the detector using the GEANT3 transport package~\cite{Brun:1994aa}. The conditions of all the ALICE detectors in terms of active channels, gain, noise level, and alignment, and their evolution with time during the data taking period, were taken into account in the simulations.

%% file: analysis.tex
\section{Analysis technique}
\label{sec:analysis}

$\Ds$ mesons and their charge conjugates were reconstructed via the decay channel $\DstophipitoKKpi$  with branching ratio $\mathrm{BR} = (2.24 \pm 0.08) \%$~\cite{Zyla:2020zbs}. The analysis was based on the reconstruction of decay-vertex topologies displaced from the interaction vertex. The separation induced by the weak decays of prompt $\Ds$ mesons is typically a few hundred of $\mum$, $c\tau \simeq 151~\mum$~\cite{Zyla:2020zbs}.

$\Ds$-meson candidates were built combining triplets of tracks with the proper charge signs, each with $|\eta| < 0.8$, at least 70 out of 159 crossed TPC pad rows, a fit quality $\chi^{2}/{\rm ndf} < 1.25$ in the TPC (where ndf is the number of degrees of freedom involved in the track fit procedure), and a minimum of two (out of six) hits in the ITS, with at least one in either of the two innermost layers, which provide the best pointing resolution. Moreover, at least 50 clusters
available for particle identification (PID) in the TPC were required and only tracks with $\pt$ above $0.6~(0.4)~\gev/c$ were considered for central (semicentral) collisions. 
These track-selection criteria limit the $\Ds$-meson acceptance in rapidity, which drops steeply to zero for $|y|>0.5$ at low $\pt$ and for $|y|>0.8$ at $\pt>5~\gevc$. Thus, only $\Ds$-meson candidates within a $\pt{}$-dependent fiducial acceptance region, $|y| < y_{\rm fid}(\pt)$, were selected. The $y_{\mathrm{fid}}(\pt)$ value was defined as a second-order polynomial function, increasing from 0.5 to 0.8 in the transverse-momentum range $0 < \pt < 5~\gevc$, and as a constant term, $y_{\mathrm{fid}}=0.8$, for $\pt > 5~\gevc$.

Unlike previous D-meson analyses based on linear selections~\cite{Acharya:2018hre,Acharya:2020pnh,RaaD}, a machine-learning (ML) approach based on Boosted Decision Trees (BDT) was adopted for the candidate selection to reduce the large combinatorial background~\cite{hipe4ml}.\ In particular, the implementation of the BDT algorithm provided by the XGBoost~\cite{Chen:2016XST} library was employed. Signal samples of prompt $\Ds$ mesons for the BDT training were obtained from Monte Carlo simulations as  described in Section~\ref{sec:detector}. The background samples were obtained from the sidebands of the candidate invariant-mass distributions in the data. Before the training, loose kinematic and topological selections were applied to the $\Ds$-meson candidates together with the PID of decay-product tracks. Pions and kaons were selected by requiring compatibility with the respective particle hypothesis within three times the detector resolution between the measured and the expected signals for either the TPC $\mathrm{d}E/\mathrm{d}x$ or the time of flight. Tracks without TOF hits were identified using only the TPC information. In addition, the absolute difference between the reconstructed ${\rm K^+K^-}$ invariant mass and the PDG average mass for the $\upphi$ meson~\cite{Zyla:2020zbs} ($\Delta M_{\mathrm{KK}}$) was required to be below $15~\mevcsquared$.
The candidate information provided to the BDTs, as an input for the models to distinguish among prompt $\Ds$ mesons and background candidates, was mainly based on the displacement of the tracks from the primary vertex, the distance between the $\Ds$-meson decay vertex and the primary vertex, the $\Ds$-meson impact parameter, and the cosine of the pointing angle between the $\Ds$-meson candidate line of flight (the vector connecting the primary and secondary vertex) and its reconstructed momentum vector. The value of $\Delta M_{\mathrm{KK}}$ and additional variables related to the PID of decay tracks were also included. 
Independent BDTs were trained in the different $\pt$ intervals of the analysis and for the different centrality intervals.\ Subsequently, they were applied to the real data sample in which the belonging class, i.e., prompt $\Ds$ meson or combinatorial background, of particle candidates is unknown. Selections on the BDT output, which is related to the candidate's probability to be a prompt $\Ds$ meson, were optimised to reject a large fraction of the combinatorial background while maintaining high signal-selection efficiency.

\subsection{Nuclear modification factor measurement}
\label{sec:analysis_raa}

The raw yields of $\Ds$ mesons, including both particles and antiparticles, were extracted from binned maximum-likelihood fits to the invariant-mass distributions. The raw yields could be extracted in transverse-momentum intervals in the ranges $2<\pt<50~\gev/c$  and $2<\pt<36~\gev/c$ for the 0--10\% and the 30--50\% centrality intervals, respectively. The fit function was composed of a Gaussian for the description of the signal and of an exponential term for the background. An additional Gaussian was used to describe the peak due to the decay $\DplustoKKpi$, with a branching ratio of $(9.68\pm 0.18)\times 10^{-3}$~\cite{Zyla:2020zbs}, present at a lower invariant-mass value than the $\Ds$-meson signal peak. 
The statistical significance of the observed signals $S/\sqrt{S+B}$, where $S$ is the raw signal
yield obtained by integrating the Gaussian function and $B$ is the background under the peak within 3 standard deviations, varies from 4 to 24 depending on the $\pt$ and centrality intervals.

The $\pt$-differential corrected yield of prompt $\Ds$ mesons was computed for each $\pt$ interval according to
\begin{equation}
  \label{eq:dNdpt}
  \left.\frac{{\rm d} N}{{\rm d}\pt}\right|_{|y|<0.5}= \frac{1}{2}\times
  \frac{1}{\Delta \pt}\times\frac{\left.\fprompt(\pt)\times N^{\rm D+\overline D,raw}(\pt)\right|_{|y|<y_{\rm fid}(\pt)}}{c_{\Delta y}(\pt)\times\effP{}(\pt)\times\mathrm{BR}\times N_\mathrm{evt}}\,.
\end{equation}
The raw-yield values $N^{\rm D+\overline D,raw}$, which contain the contribution of non-prompt $\Ds$ mesons from beauty-hadron decays, were divided by a factor of two and multiplied by the prompt fraction $\fprompt$ to obtain the charge-averaged yields of prompt $\Ds$ mesons. Furthermore, they were divided by the acceptance times efficiency of prompt $\Ds$ mesons $\effP{}$, the BR of the decay channel, the width of the $\pt$ interval $\Delta \pt$, the correction factor for the rapidity coverage $c_{\Delta y}$, and the number of analysed events $N_{\rm evt}$. The correction factor for the rapidity acceptance $c_{\Delta y}$ was computed with FONLL perturbative QCD calculations~\cite{Cacciari:1998it,Cacciari:2001td}. It was defined as the ratio between the generated D-meson yield in $\Delta y = 2\,y_{\rm fid}$ and that in $|y|<0.5$. The resulting values were in agreement within 1\% with PYTHIA~8 simulations for pp collisions. To account for possible differences in Pb--Pb collisions and as an extreme variation, a flat rapidity distribution was also considered. The discrepancies with respect to FONLL calculations were negligible in comparison to other sources of systematic uncertainty described in Section~\ref{sec:syst}.

\begin{figure}[tb]
    \begin{center}
    \includegraphics[width = 0.48\textwidth]{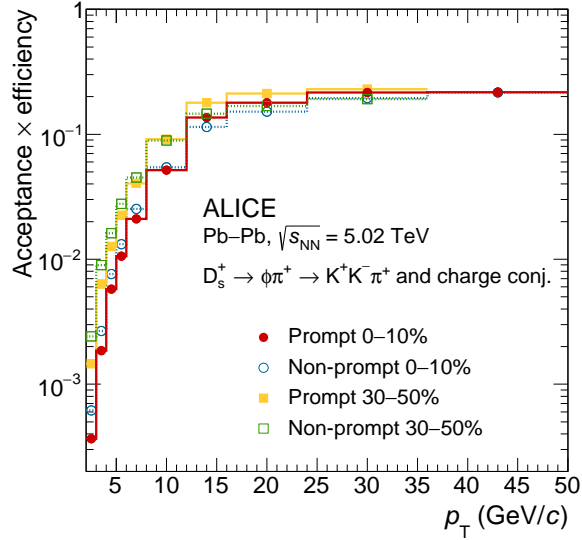}
    \end{center}
    \caption{Acceptance-times-efficiency factor for $\Ds$ mesons as a function of $\pt$. The $(\rm Acc \times \epsilon)$ factors for prompt (red) and non-prompt (blue) $\Ds{}$ mesons in Pb--Pb collisions for the 0--10\% centrality interval are shown, together with those for prompt (orange) and non-prompt (green) $\Ds{}$ mesons for the 30--50\% centrality interval.}
    \label{fig:eff}
\end{figure}

The $(\rm Acc \times \epsilon)$ correction was obtained from the simulations described in Section~\ref{sec:detector} using samples not employed in the BDT training.
The $\Ds{}$-meson $\pt{}$ distributions from simulations were reweighted in order to use realistic momentum distributions in the determination of the $(\rm Acc \times \epsilon)$ factor, which depends on $\pt{}$. In particular, the weights were defined to match the shape given by FONLL calculations multiplied by the $\RAA{}$ of $\Ds$ mesons predicted by the TAMU~\cite{He:2019vgs} model.
The $(\rm Acc \times \epsilon)$ factors as a function of $\pt$ for prompt and non-prompt $\Ds$ mesons in the 0--10\% and 30--50\% centrality intervals are shown in Fig.~\ref{fig:eff}. The difference between the $(\rm Acc \times \epsilon)$ factor for prompt and non-prompt $\Ds{}$ mesons arises from the BDT selections applied, given the different decay topology of $\Ds{}$ mesons coming from beauty-hadron decays. In particular, the non-prompt $\Ds{}$ mesons are on average more displaced from the primary vertex due to the large beauty-hadron lifetime, $c\tau \simeq 500~\mum$~\cite{Zyla:2020zbs}, and therefore are more efficiently selected in the low-$\pt{}$ region. At high $\pt{}$, where the candidate decay length is less important to separate signal from background, the BDT selections are able to suppress the non-prompt efficiency with respect to the prompt one. The $(\rm Acc \times \epsilon)$ is higher for semicentral collisions, by up to a factor two at low $\pt{}$, since less stringent selections can be applied thanks to the lower combinatorial background.

The $\fprompt$ fraction in each $\pt{}$ interval was obtained following the procedure employed in Refs.~\cite{Adam:2015sza,Acharya:2018hre,RaaD}. The calculation was based on the beauty-hadron production cross sections in pp collisions at $\sqrt{s} = 5.02~\mathrm{TeV}$  from FONLL calculations, the beauty hadron to $\mathrm{D+X}$ decay kinematics from the PYTHIA~8 decayer, the $(\rm Acc \times \epsilon)$ correction factor for non-prompt $\Ds$ mesons, and the $\av{\TAA}$ for the corresponding centrality interval. In addition, the nuclear modification factor of $\Ds$ mesons from beauty-hadron decays was accounted for and $R_\mathrm{AA}^{\mathrm{prompt}} = R_\mathrm{AA}^{\mathrm{non\text{-}prompt}}$ was assumed as in Ref.~\cite{Acharya:2018hre}. The values of $\fprompt$ range between 0.86 and 0.91 depending on the $\pt$ interval and the centrality interval.

The prompt $\Ds{}$-meson nuclear modification $\RAA{}$ factor was computed following Eq.~\ref{eq:Raa}. The measurement of the $\pt{}$-differential cross section of prompt $\Ds{}$ mesons with $|y| < 0.5$ in pp collisions at $\sqrt{s} = 5.02~\mathrm{TeV}$ from Ref.~\cite{Acharya:2021cqv}, which reaches up to $\pt{} = 24~\gevc{}$, was used as a reference for the $\RAA{}$ computation. At higher $\Ds{}$-meson $\pt{}$, $24 < \pt{} < 50~\gevc{}$, FONLL calculations were used as a reference by scaling the predictions to match the measured values at lower $\pt{}$. The $\pt{}$-extrapolation procedure is the same as in Ref.~\cite{Adam:2015sza}. As an example, the total systematic uncertainty of the pp reference in the $36 < \pt{} < 50~\gevc{}$ interval is $^{+42}_{-33}\%$.

\subsection{Elliptic flow measurement}
\label{sec:analysis_v2}
The elliptic flow of prompt $\Ds{}$ mesons was measured for semicentral events in transverse-momentum intervals in the range $2<\pt<24~\gev/c$. The same ML models trained for the $\RAA{}$ measurement in the 30--50\% centrality interval were used and the same selections on the BDT output were applied.
The analysis procedure for the $\vtwo$ determination followed closely with what was done in Ref.~\cite{Acharya:2020pnh} for the measurement of the non-strange D-meson elliptic flow.
The $\Ds{}$-meson $\vtwo$ coefficients were measured using the Scalar Product (SP) method~\cite{Voloshin:2008dg, Luzum:2012da} and can be expressed as
\begin{equation}
\label{eq:SP}
\vtwo\{\mathrm{SP}\}=\langle\langle \pmb{u}_{\mathrm{2}}\cdot\frac{\pmb{Q}_{\mathrm{2}}^{\rm A*}}{M^{\rm A}}\rangle\rangle\bigg{/}\sqrt{\frac{\langle\frac{\pmb{Q}_{\mathrm{2}}^{\rm A}}{M^{\rm A}}\cdot\frac{\pmb{Q}_{\mathrm{2}}^{\rm B*}}{M^{\rm B}}\rangle\langle\frac{\pmb{Q}_{\mathrm{2}}^{\rm A}}{M^{\rm A}}\cdot\frac{\pmb{Q}_{\mathrm{2}}^{\rm C*}}{M^{\rm C}}\rangle}{\langle\frac{\pmb{Q}_{\mathrm{2}}^{\rm B}}{M^{\rm B}}\cdot\frac{\pmb{Q}_{\mathrm{2}}^{\rm C*}}{M^{\rm C}}\rangle}},
\end{equation} 
where \textbf{\textit{u}}$_2 = e^{i2\varphi_\mathrm{D}}$ is the unit flow vector of the D-meson candidate with azimuthal angle $\varphi_{\mathrm{D}}$, $\pmb{Q}^k_2$ is the subevent $2^\mathrm{nd}$-harmonic flow vector for the subevent $k$, and $M^k$ represents the subevent multiplicity. The SP denominator was calculated with the formula introduced in Ref.~\cite{Luzum:2012da}, where the three subevents, indicated as A, B, and C, are defined by the particles measured in the V0C, V0A, and TPC detectors, respectively. For the TPC detector, the $\pmb{Q}_2$ vector was computed from the azimuthal angles  of charged tracks reconstructed with $|\eta|<0.8$ and $M$ was the number of measured tracks. For the V0A and V0C detectors, the $\pmb{Q}_2$ vectors were calculated from the azimuthal distribution of the energy deposition in the detector sectors and $M$ was the sum of the amplitudes measured in each channel~\cite{Acharya:2020pnh}. The $\pmb{Q}_2$ vectors were corrected for detector effects arising from the non-uniform acceptance~\cite{Selyuzhenkov:2007zi}. The single bracket $\langle\rangle$ in Eq.~\ref{eq:SP} refers to an average over all the events, while the double brackets $\langle\langle\rangle\rangle$ denote the average over all particles in the considered $\pt$ interval and all events. The SP denominator was obtained as a function of the collision centrality.

\begin{figure}[tb]
    \begin{center}
    \includegraphics[width = 0.4\textwidth]{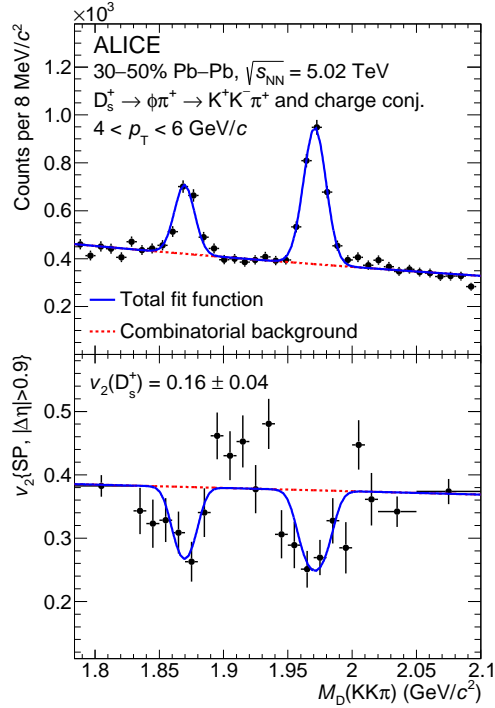}
    \end{center}
    \caption{Simultaneous fit to the invariant-mass spectrum and $v_2(M_{\mathrm{D}})$ of $\Ds{}$-meson candidates in the $4 < \pt{} < 6~\gevc{}$ interval for the 30--50\% centrality interval. The solid blue and the dotted red curves represent the total and combinatorial-background fit functions, respectively.}
    \label{fig:inv_mass}
\end{figure}

The elliptic flow of $\Ds{}$ mesons cannot be directly measured using Eq.~\ref{eq:SP} as signal candidates cannot be identified on a particle-by-particle basis. The measured anisotropic flow coefficient $\vtwotot$ can be written as a weighted sum of the $\vtwo$ of candidates reconstructed from true $\Ds{}$-meson decays ($\vtwosig$) and that of the background ($\vtwobkg$)~\cite{Borghini:2004ra}
\begin{equation}
\label{eq:vtwoTot}
	\vtwotot (M_\mathrm{D}) = \frac{1}{\Nsig + \Nbkg + \Ndplus}(M_\mathrm{D})\bigg[\Nsig (M_\mathrm{D})\vtwosig + \Nbkg (M_\mathrm{D})\vtwobkg (M_\mathrm{D}) + \Ndplus (M_\mathrm{D})\vtwodplus\bigg],
\end{equation}
where $\Nsig$ and $\Nbkg$ are the raw signal and background yields, respectively. An additional $\vtwodplus$ free parameter and the corresponding raw yield $\Ndplus$ were included to account for the $\DplustoKKpi$ contribution to the measured $\vtwotot$ distribution.
A simultaneous fit to the invariant-mass spectrum and the $\vtwotot$ distribution as a function of the invariant mass was performed in each $\pt$ interval to extract the elliptic flow coefficients. The fit function for the invariant-mass distributions was composed of two Gaussian terms to describe the signal and the peak due to the decay $\DplustoKKpi$, and an exponential distribution for the background, as for the $\RAA{}$ measurement of Section~\ref{sec:analysis_raa}.
The $\vtwosig$ was measured from a fit to the $\vtwotot$ distribution with the function of Eq.~\ref{eq:vtwoTot}, where $\vtwobkg(M_\mathrm{D})$ was described by a linear function. 
Figure~\ref{fig:inv_mass} shows the simultaneous fit to the invariant-mass spectrum and $v_2^\mathrm{tot}(M_\mathrm{D})$ of $\Ds{}$ mesons in the $4 < \pt{} < 6~\gevc{}$ interval for the 30--50\% centrality interval.

The reconstructed $\Ds{}$-meson signal is a mixture of prompt $\Ds{}$ mesons and non-prompt $\Ds{}$ mesons from beauty-hadron decays. Therefore, the $\vtwosig$ can be expressed as a linear combination of prompt ($\vtwoprompt$) and non-prompt ($\vtwofd$) contributions weighted by the fraction of prompt ($\fprompt$) and non-prompt $(1 - \fprompt{})$ $\Ds{}$ mesons in the extracted signal, respectively. 
The fraction of promptly produced $\Ds{}$ mesons was estimated as a function of $\pt$ with the theory-driven method described in Section~\ref{sec:analysis_raa}. The $\vtwo{}$ coefficients of prompt $\Ds{}$ mesons were obtained assuming $\vtwofd=\vtwoprompt/2$. This hypothesis is based on the $\vtwo{}$ measurements of the non-prompt $\Jpsi$ performed by ATLAS and CMS~\cite{ATLAS:2018xms,Khachatryan:2016ypw}, and on the available model calculations~\cite{Uphoff:2012gb,Aichelin:2012ww,Greco:2007sz} that indicate $0< \vtwofd < \vtwoprompt$.

%% file: systematics.tex
\section{Systematic uncertainties}
\label{sec:syst}

\subsection{Nuclear modification factor measurement}
\label{sec:syst_raa}
The measurement of the $\Ds{}$-meson corrected yield is affected by the following sources of systematic uncertainties: (i) the raw-yield extraction from the invariant-mass distributions, (ii) the track-reconstruction efficiency, (iii) the PID and selection efficiency, (iv) the generated $\Ds{}$-meson $\pt{}$ shape in the simulation, and (v) the prompt fraction estimation. 
In addition, the uncertainty due to the branching ratio of 3.6\%~\cite{Zyla:2020zbs}, and that due to the centrality-interval definition were considered.  This last contribution arises from the uncertainty of the fraction of the hadronic cross section used in the Glauber fit to determine the centrality, and was estimated to be $<0.1$\% and 2\% for the 0--10\% and 30--50\% centrality intervals, respectively~\cite{Adam:2015sza}.
A procedure similar to that described in Refs.~\cite{RaaD,Acharya:2018hre} was used to estimate the uncertainties as a function of the $\pt{}$ interval and the centrality interval. The estimated values of the systematic uncertainties are summarised in Table~\ref{tab:systunc} for representative $\pt{}$ intervals, together with the total systematic uncertainty obtained from the sum in quadrature of the different contributions.

\begin{table}[tb]
\caption{Relative systematic uncertainties of the prompt $\Ds{}$-meson corrected yield in Pb–Pb collisions for central and semicentral events in representative $\pt{}$ intervals.}
\centering
\renewcommand*{\arraystretch}{1.4}
\begin{tabular}[t]{l|>{\centering}p{0.05\linewidth}>{\centering}p{0.05\linewidth}|>{\centering}p{0.05\linewidth}>{\centering}p{0.05\linewidth}}
\toprule
Centrality interval & \multicolumn{2}{c|}{0--10\%} & \multicolumn{2}{c}{30--50\%} \\
$\pt~(\GeV/c)$ & 2--3 & \multicolumn{1}{c|}{12--16} & 2--3 & \multicolumn{1}{c}{12--16} \\
\midrule
Yield extraction & 8\% & \multicolumn{1}{c|}{2\%} & 8\% & \multicolumn{1}{c}{3\%} \\
Tracking efficiency & 12\% & \multicolumn{1}{c|}{12\%} & 10\% & \multicolumn{1}{c}{8\%} \\
Selection efficiency & 9\% & \multicolumn{1}{c|}{4\%} & 5\% & \multicolumn{1}{c}{3\%} \\
Prompt fraction & $^{+8}_{-16}$\% & \multicolumn{1}{c|}{$^{+9}_{-18}$\%} & $^{+8}_{-16}$\% & \multicolumn{1}{c}{$^{+8}_{-17}$\%} \\
MC $\pt$ shape & 5\%   & \multicolumn{1}{c|}{negl.} & 3\% & \multicolumn{1}{c}{negl.} \\
Centrality limits & \multicolumn{2}{c|}{$<0.1$\%} & \multicolumn{2}{c}{$2\%$} \\
\midrule
Branching ratio & \multicolumn{4}{c}{4\%} \\
\midrule
Total syst. unc. & $^{+20}_{-24}$\% & \multicolumn{1}{c|}{$^{+16}_{-23}$\%} & $^{+17}_{-22}$\% & \multicolumn{1}{c}{$^{+13}_{-20}$\%} \\
\bottomrule
\end{tabular}
\label{tab:systunc}	
\end{table}

The systematic uncertainty of the raw-yield extraction was evaluated by repeating the fit of the invariant-
mass distribution varying the lower and upper limits of the fit range, the bin width, and the functional form of the background fit function.
The systematic uncertainty was defined as the RMS of the distribution of the signal yields obtained from all these variations and ranges from 2\% to 8\% depending on the centrality interval and the $\pt{}$ interval.

The systematic uncertainty of the track-reconstruction efficiency was estimated by varying the track-
quality selection criteria and by comparing the prolongation probability of the TPC tracks to the ITS
hits in data and simulation. The comparison was performed after weighting the relative abundances of primary and secondary particles in the simulation to match those observed in data~\cite{ALICE-PUBLIC-2017-005}. The estimated
uncertainty ranges from 5\% to 14\%.

The systematic uncertainty of the selection efficiency originates from imperfections in the description of the detector resolutions and alignments in the simulation. It was estimated by comparing the corrected yields obtained by repeating the analysis with different selections on the BDT output, which resulted in up to 50\% higher and lower efficiencies with respect to the central values. The assigned systematic uncertainty ranges from 3\% to 9\%.
Possible systematic effects due to the loose PID selection, applied prior to the machine-learning one, were investigated comparing  pion and kaon PID selection efficiencies in data and in simulations. A pure sample of pions was selected from $\kzero$ and $\lmb$ decays, while samples of kaons in the TPC (TOF) were obtained applying a strict PID selection using the TOF (TPC) information. Since no significant differences were observed, no systematic uncertainty was assigned.

An additional contribution to the systematic uncertainty of the efficiency originates from possible differences between the real and simulated $\Ds{}$-meson $\pt{}$ distributions. It was estimated by calculating the efficiency using alternative $\Ds{}$-meson $\pt$ shapes obtained by re-weighting the $\pt$ distribution from MC simulations to match those predicted by theoretical models. The $\pt{}$ distributions from FONLL calculations including or not hot-medium effects, parametrised using the $\pt{}$-differential $\RAA{}$ from the LGR~\cite{Li:2019lex}, PHSD~\cite{Song:2015ykw}, TAMU~\cite{He:2019vgs}, and Catania~\cite{Scardina:2017ipo} models, were considered. The resulting uncertainty was estimated to be about 5\% and 3\% for the 0--10\% and 30--50\% centrality intervals, respectively, in the lowest $\pt{}$ intervals where the efficiency varies steeply with $\pt{}$, and to decrease to zero above $12~\GeV{}/c$.

The systematic uncertainty of the prompt fraction was estimated by varying the FONLL parameters (b-quark mass, factorisation, and renormalisation scales, according to the prescription reported in Ref.~\cite{Cacciari:2012ny}) in the calculation of the $\pt{}$-differential production cross section of non-prompt $\Ds{}$ mesons. In addition, the ratio of the non-prompt and prompt $\Ds{}$-meson $\RAA{}$ was varied in the range $\frac{1}{3} < R_\mathrm{AA}^{\mathrm{non\text{-}prompt}}/R_\mathrm{AA}^{\mathrm{prompt}}<3$ as done in Ref.~\cite{Acharya:2018hre}. The resulting uncertainty ranges between $^{+8}_{-16}$\% and $^{+12}_{-23}$\%.

In the $\RAA{}$ calculation, the BR uncertainty of the $\Ds{}$-meson yield in Pb--Pb collisions and of the pp reference cross section cancels out in the ratio. The contribution due to the prompt fraction uncertainty, estimated by the variation of the parameters of the FONLL calculation, was considered to be fully correlated and the remaining systematic uncertainties were propagated as uncorrelated.
The uncertainties of the $\RAA{}$ normalisation are the quadratic sum of the pp normalisation uncertainty, 2.1\%~\cite{ALICE-PUBLIC-2018-014}, the $\av{\TAA{}}$ uncertainty, 0.7\%  (1.5\%) for the 0--10\% (30--50\%) centrality interval~\cite{ALICE-PUBLIC-2018-011}, and the one related to the centrality-interval definition described above.

\subsection{Elliptic flow measurement}
\label{sec:syst_v2}
The systematic uncertainties of the measurement of the $\Ds{}$-meson $\vtwo$ coefficients were estimated with procedures similar to those detailed in Ref.~\cite{Acharya:2020pnh}. They include the following sources: (i) the signal extraction from the invariant-mass and $\vtwotot$ distributions, (ii) the non-prompt $\Ds{}$ contribution, and (iii) the centrality dependence of the SP denominator. The selection efficiency was observed to be independent of the $\Ds{}$-meson azimuthal direction, therefore no contribution to the systematic uncertainty was assigned. The non-flow effects are naturally suppressed due to the pseudorapidity gap of at least 0.9 units between the pseudorapidity interval used for the $\Ds$-meson reconstruction, and the V0C used for the $\pmb{Q}_2$-vector determination. The estimated values of the systematic uncertainties are summarised in Table~\ref{tab:systuncv2} for representative $\pt{}$ intervals.

\begin{table}[tb]
\caption{Systematic uncertainties of the prompt $\Ds{}$-meson $\vtwo{}$ in Pb–Pb collisions for the 30--50\% centrality interval in representative $\pt{}$ intervals. The uncertainties of the fitting procedure and non-prompt contribution subtraction are quoted as absolute uncertainties, while that of the SP denominator as relative uncertainty.}
\centering
\renewcommand*{\arraystretch}{1.4}
\begin{tabular}[t]{l|>{\centering}p{0.05\linewidth}>{\centering}p{0.05\linewidth}}
\toprule
$\pt~(\GeV/c)$ & 2--4 & \multicolumn{1}{c}{12--16} \\
\midrule
$M$ and $\vtwo$ fits & 0.01 & \multicolumn{1}{c}{0.02} \\
Non-prompt contribution & $^{+0.031}_{-0.007}$ & \multicolumn{1}{c}{$^{+0.028}_{-0.006}$} \\
SP denominator & \multicolumn{2}{c}{0.5\%} \\
\bottomrule
\end{tabular}
\label{tab:systuncv2}	
\end{table}

The uncertainty due to the simultaneous fit was estimated by repeating the fit several times with different configurations, as done for the $\RAA$ measurement. 
The RMS of the $\vtwo$ distribution obtained from the different trials, separately for each $\pt$ interval, was assigned as systematic uncertainty. The absolute systematic uncertainty values due to the signal extraction range between 0.01 and 0.03 depending on $\pt{}$.

The systematic uncertainty related to the correction for the contribution of non-prompt $\Ds{}$ to the measured $\vtwo{}$ has two main sources. The first one is due to the $\fprompt$ calculation and it was treated as described in Section~\ref{sec:syst_raa} for the $\RAA{}$ measurement. The second source is due to the assumption of $\vtwofd$ = $\vtwoprompt/2$. This was estimated by considering a flat distribution of $\vtwofd$ between 0 and $\vtwoprompt$ and by varying the central value of $\vtwofd$ by $\pm\vtwoprompt/\sqrt{12}$, corresponding to one standard deviation. The values of the absolute systematic uncertainty from the non-prompt correction range between $^{+0.020}_{-0.005}$ and $^{+0.039}_{-0.009}$ for the different $\pt{}$ intervals.

The contribution of the SP denominator to the systematic uncertainty is due to the centrality dependence. The uncertainty was evaluated as the difference of the centrality integrated value, computed from the events in the 30--50\% interval, with that obtained as weighted average of SP-denominator values in narrow centrality intervals using the $\Ds{}$-meson yields as weights. A systematic uncertainty of 0.5$\%$ was assigned.

%% file: results.tex
\section{Results}

\begin{figure}[!tb]
    \begin{center}
    \includegraphics[width = 0.48\textwidth]{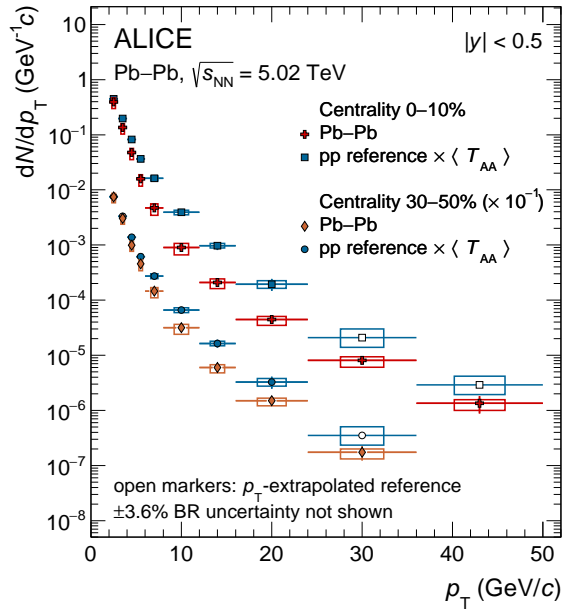}
    \end{center}
    \caption{$\pt$-differential production yields of prompt $\Ds$ mesons in the 0--10\% and 30--50\% centrality intervals in Pb--Pb collisions at $\sqrtsNN=5.02~\TeV$ compared to the pp reference~\cite{Acharya:2021cqv} scaled by the average nuclear overlap function $\av{\TAA}$ of the corresponding centrality interval. The open markers indicate where the pp reference is extrapolated using FONLL calculations. The $\pt$-differential yields in the 30--50\% centrality interval and the corresponding pp reference are scaled by a factor of $10^{-1}$ for better visibility. Statistical uncertainties (bars) and systematic uncertainties (boxes) are shown.}
    \label{fig:pt_diff_yields}
\end{figure}

The $\pt$-differential production yields $\de N/\de\pt$ of prompt $\Ds$ mesons measured in the 0--10\% and 30--50\% centrality intervals are shown in Fig.~\ref{fig:pt_diff_yields}. For the semicentral class of events, the measurements are scaled by $10^{-1}$ for better visibility. The results are compared with the pp reference cross section multiplied by the corresponding average nuclear overlap function $\av{\TAA}$. The larger data sample and the improved analysis technique enable an extended $\pt$ coverage and finer $\pt$ intervals in the measured $\de N/\de\pt$ of prompt $\Ds$ mesons compared to the previous measurement by the ALICE Collaboration in Pb--Pb collisions at $\sqrtsNN=5.02~\TeV$~\cite{Acharya:2018hre}. A strong suppression of the $\Ds$ yields compared to the binary-scaled pp reference is observed for both centrality intervals for $\pt > 3\text{--}4~\GeV/c$, similarly as for the non-strange D mesons~\cite{RaaD}. This suppression is understood in terms of modification of the charm-quark momentum spectra due to the interactions within the QGP.

\begin{figure}[!tb]
    \begin{center}
    \includegraphics[width = 0.9\textwidth]{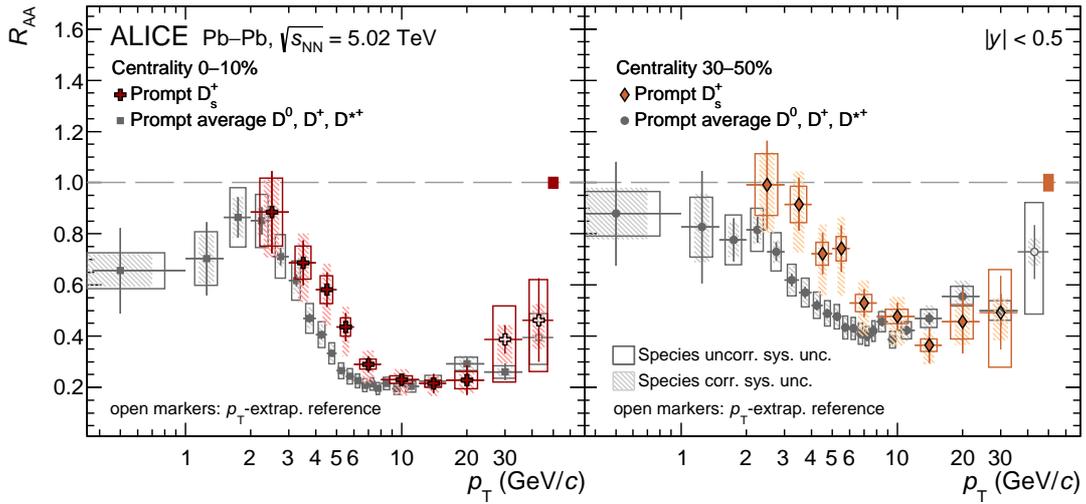}
    \end{center}
    \caption{Nuclear modification factor $\RAA$ of prompt $\Ds$ mesons in the 0--10\% (left panel) and 30--50\% (right panel) centrality intervals in Pb--Pb collisions at $\sqrtsNN=5.02~\TeV$ compared with the one of prompt non-strange D mesons (average of $\Dzero$, $\Dplus$, and $\Dstar$)~\cite{RaaD}. The empty (filled) boxes represent the species uncorrelated (correlated) systematic uncertainties. The normalisation uncertainty is represented by a filled box at $\RAA=1$.}
    \label{fig:RAA}
\end{figure}
The nuclear modification factor $\RAA$ of prompt $\Ds$ mesons is compared with the average $\RAA$ of prompt $\Dzero$, $\Dplus$, and $\Dstar$ mesons in Fig.~\ref{fig:RAA} for the 0--10\% and 30--50\% centrality intervals, in the left and right panels, respectively. 
The systematic uncertainties related to the tracking efficiency and the prompt-fraction estimation are considered as fully correlated between the different D-meson species, and are reported separately from the other sources of systematic uncertainty which are uncorrelated.
The $\RAA$ of $\Ds$ and non-strange D mesons show a minimum value of about 0.2 (0.4) around $\pt\approx 10~\GeV/c$ in the 0--10\% (30--50\%) centrality interval. For lower $\pt$, the $\RAA$ increases with decreasing $\pt$ reaching about unity around $\pt \approx 2\text{--}3~\GeV/c$. In both centrality intervals, the $\RAA$ of prompt $\Ds$ and non-strange D mesons are compatible within uncertainties for $\pt \gtrsim 10~\GeV/c$. In this $\pt$ region, the hadronisation is expected to occur mainly via fragmentation and the dominant effect leading to the observed suppression is the charm-quark energy loss in the QGP. For lower $\pt$, the measured $\RAA$ of prompt $\Ds$ mesons is systematically higher than that of non-strange D mesons but compatible within about one standard deviation of the combined statistical and systematic uncertainties.

\begin{figure}[!tb]
    \begin{center}
    \includegraphics[width = 0.9\textwidth]{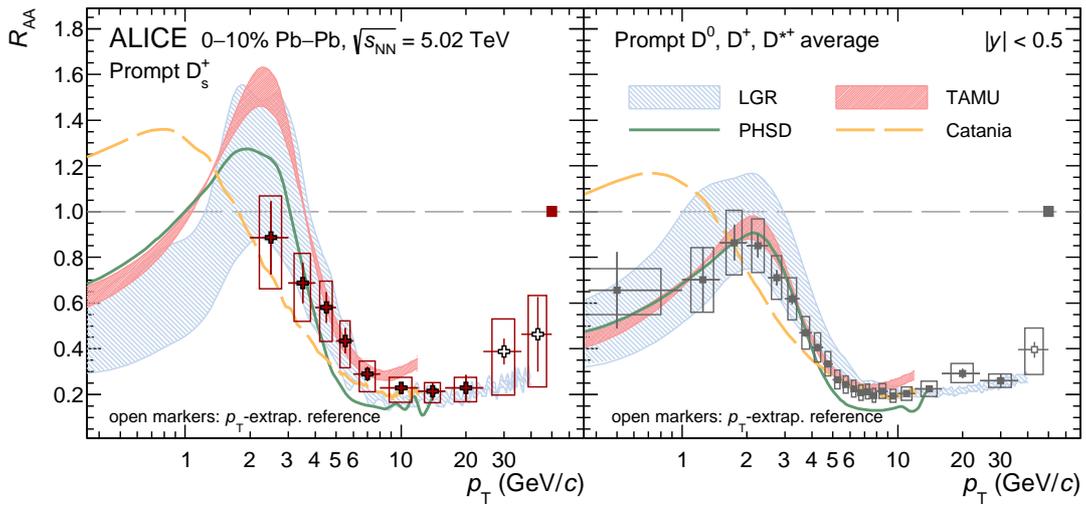}
    \end{center}
    \caption{Nuclear modification factor $\RAA$ of prompt $\Ds$ mesons (left panel) and non-strange D mesons~\cite{RaaD} (right panel) in the 0--10\% centrality interval in Pb--Pb collisions at $\sqrtsNN=5.02~\TeV$ compared with theoretical calculations based on charm-quark transport in a hydrodynamically expanding QGP implementing strangeness enhancement and hadronisation of charm quarks via coalescence in addition to fragmentation in the vacuum~\cite{Song:2015sfa, Song:2015ykw, He:2019vgs, Li:2019lex, Scardina:2017ipo}. The boxes represent the total systematic uncertainties. The colour bands represent the theoretical uncertainty when available.}
    \label{fig:RAAvsModels}
\end{figure}

In the left and right panels of Fig.~\ref{fig:RAAvsModels}, the $\RAA$ of prompt $\Ds$ and non-strange D mesons in the 0--10\% centrality interval are compared with theoretical calculations implementing charm-quark transport in the QGP~\cite{Berrehrah:2016vzw}.
All the models include an enhancement of the strangeness content of the QGP and the hadronisation of charm quarks is implemented either via fragmentation, which is dominant at high $\pt$, or via coalescence with light quarks in the QGP. In the Catania~\cite{Scardina:2017ipo,Plumari:2017ntm} and LGR~\cite{Li:2019lex} models the coalescence occurs instantaneously at the phase boundary and is implemented through the Wigner formalism~\cite{Dover:1991zn}. In the PHSD model~\cite{Song:2015sfa, Song:2015ykw}, the hadronisation in heavy-ion collisions is described via a Monte Carlo simulation of the coalescence process in competition to fragmentation.
In the TAMU~\cite{He:2019vgs} model, the hadronisation via coalescence proceeds via formation of resonant states when approaching the (pseudo)critical temperature within the formalism of a Resonance Recombination Model~\cite{Ravagli:2007xx}. 
For the description of the D-meson $\pt$ spectra in pp collisions, all the models use as starting point FONLL calculations~\cite{Cacciari:1998it,Cacciari:2001td,Cacciari:2012ny}. Charm quarks are hadronised in pp collisions with fragmentation in the PHSD and LGR models, while in the Catania model the charm-quark hadronisation via coalescence is also implemented in addition to that via fragmentation~\cite{Minissale:2020bif}. In pp collisions, the hadronisation in the TAMU model is instead determined with a statistical hadronisation approach, in which the strangeness production is suppressed in pp with respect to heavy-ion collisions. This is described with a suppression factor for strange particles of $\gamma_\mathrm{s}=0.6$~\cite{He:2019tik}, which is instead unity in heavy-ion collisions.
All the models reproduce qualitatively the measured $\RAA$ of prompt $\Ds$ and non-strange D mesons. The Catania model underestimates both measurements for $2<\pt<5~\GeV/c$ by about $2\,\sigma$ of the combined statistical and systematic uncertainties of the measured points, while it overestimates the non-strange D-meson $\RAA$ for $\pt<1.5~\GeV/c$, where no measurement is available for strange mesons. In contrast, the PHSD model describes well the measured nuclear modification factors for $\pt<5~\GeV/c$ and underestimates them by about $2\,\sigma$ for higher $\pt$. The TAMU model describes the measurements within uncertainties, with a tension of about $2\,\sigma$ of the combined statistical and systematic uncertainties of the $\Ds$-meson measurement in $2<\pt<3~\GeV/c$. These three models do not include charm-quark interactions with medium constituents via radiative processes, hence are not expected to describe the $\RAA$ of strange and non-strange D mesons for $\pt>6\text{--}8~\GeV/c$. The LGR model, which instead includes gluon-radiation processes, provides a good description of the $\RAA$ up to high $\pt$. All the models predict a smaller suppression of the $\Ds$-meson $\RAA$ compared to non-strange D mesons at low and intermediate $\pt$.

\begin{figure}[!tb]
    \begin{center}
    \includegraphics[width = 0.9\textwidth]{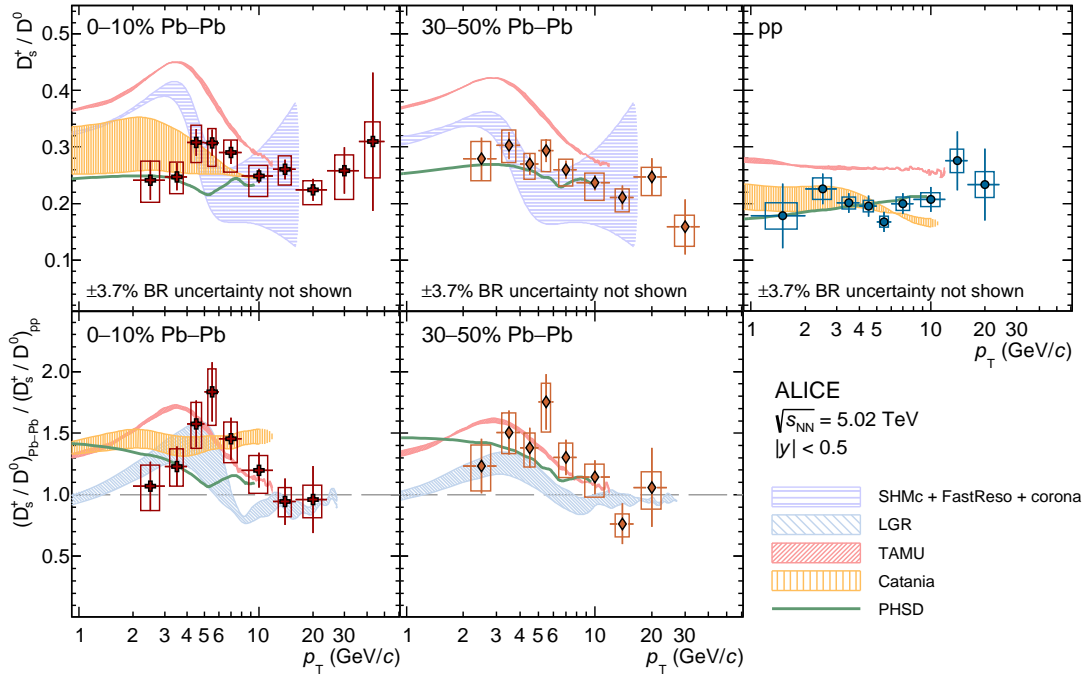}
    \end{center}
    \caption{Top panels: $\Ds/\Dzero$ $\pt$-differential production ratios in the 0--10\% (left panel) and 30--50\% (middle panel) centrality intervals in Pb--Pb collisions at $\sqrtsNN=5.02~\TeV$ and in pp collisions (right panel) at the same centre-of-mass energy compared with theoretical calculations based on charm-quark transport in a hydrodynamically expanding QGP~\cite{Song:2015sfa, Song:2015ykw, He:2019vgs, Li:2019lex, Plumari:2017ntm, He:2019tik, Minissale:2020bif} and on statistical hadronisation~\cite{Andronic:2021erx}. Bottom panels: $\Ds/\Dzero$ $\pt$-differential ratios in Pb--Pb collisions divided by those in pp collisions, in the 0--10\% (left panel) and 30--50\% (right panel) centrality intervals, compared with theoretical calculations.}
    \label{fig:DsOverD}
\end{figure}
The possible enhancement of the yield of D mesons with strange-quark content with respect to that of non-strange D mesons was further investigated by computing the ratio between the $\pt$-differential production yields of prompt $\Ds$ mesons and those of prompt $\Dzero$ mesons~\cite{RaaD}. The systematic uncertainty related to the determination of the tracking efficiency and the contribution due to the subtraction of the component from beauty-hadron decays were propagated as fully correlated in the ratios, while all the other sources of systematic uncertainties were considered as uncorrelated between the measurements of $\Ds$ and $\Dzero$ mesons. The top row of Fig.~\ref{fig:DsOverD} shows the $\Ds/\Dzero$ yield ratios in the 0--10\% (left panel) and 30--50\% (middle panel) centrality intervals compared to the same quantity measured in minimum-bias pp collisions~\cite{Acharya:2021cqv} (right panel) and to theoretical calculations. The $\Ds/\Dzero$ yield ratios in Pb--Pb collisions divided by those measured in pp collisions are shown in the bottom row of the same figure.\ The average values of the $\Ds/\Dzero$ ratios in the  $2<\pt<8~\GeV/c$  interval are higher in Pb--Pb collisions than those in pp collisions by about $2.3\sigma$ and $2.4\sigma$ of the combined statistical and systematic uncertainties, for the 0--10\% and 30--50\% centrality intervals,  respectively. In central collisions, the measured $\Ds/\Dzero$ ratio is compatible with the one measured by the STAR Collaboration in Au--Au collisions at $\sqrtsNN=200~\GeV$~\cite{Adam:2021qty}.
The $\Ds/\Dzero$ ratios in pp and in central (central and semicentral) Pb--Pb collisions are described within uncertainties by the Catania (PHSD) model. The TAMU model significantly overestimates the measured $\Ds/\Dzero$ by a similar amount in the two colliding systems, leading to a good description of the ratio of the $\Ds/\Dzero$ measured in Pb--Pb and pp collisions, as shown in the bottom panels of Fig.~\ref{fig:DsOverD}. While the Catania and PHSD models predict a $\Ds/\Dzero$ ratio almost $\pt$ independent for $\pt<3~\GeV/c$ and then mildly decreasing towards the pp value at higher $\pt$, the TAMU and LGR models predict a peak around $\pt\approx 3\text{--}4~\GeV/c$. The origin of such a peak would be motivated by the different masses of $\Ds$ and $\Dzero$ mesons and by the collective radial expansion of the system with a common flow-velocity profile, which imposes an equal velocity boost to all particles in case of complete thermalisation. In addition, also the hadronisation via coalescence is expected to modify the $\pt$ shape of the $\Ds/\Dzero$ ratio due to the different masses of u and s quarks. A similar $\pt$ shape is predicted by the GSI-Heidelberg statistical hadronisation model (SHMc)~\cite{Andronic:2021erx}, which is reported in the top panels of Fig.~\ref{fig:DsOverD} for central and semicentral Pb--Pb collisions, where the $\pt$ spectra of charm hadrons are modelled with a core-corona approach. The low-$\pt$ region is dominated by the core contribution described with a Blast Wave function. The corona contribution is instead parametrised from measurements in pp collisions and is relevant at high $\pt$. The $\pt$-spectra modification due to resonance decays is computed using the FastReso package~\cite{Mazeliauskas:2018irt}.
Within the current uncertainties of the measurement, no firm conclusions can be drawn on the $\pt$ shape of the $\Ds/\Dzero$ ratio in Pb--Pb collisions at low and intermediate $\pt$. These results however provide important indications about the role of the charm-quark hadronisation via coalescence in the QGP, complementary to those obtained via the simultaneous comparison of the measured D-meson $\RAA$ and $v_\mathrm{n}$ coefficients~\cite{RaaD,Acharya:2020pnh}.

The visible production yield of prompt $\Ds$ mesons was evaluated by integrating the $\pt$-differential yield over the narrower $\pt$ intervals of the measurement. The systematic uncertainties were propagated as fully correlated among the measured $\pt$ intervals, except for the raw-yield extraction uncertainty, which was treated as uncorrelated considering the variations of the signal-to-background ratio and the shape of the combinatorial-background distribution as a function of $\pt$. In order to obtain the $\pt$-integrated production yield, the $\de N/\de\pt$ was extrapolated in $0<\pt<2~\GeV/c$. For this purpose, the measured $\pt$-differential $\Ds/\Dzero$ ratio was interpolated using the shape predicted by the PHSD model and leaving the normalisation as a free parameter. The extrapolated $\Ds/\Dzero$ ratio for $\pt<2~\GeV/c$ was then multiplied by the $\de N/\de\pt$ of $\Dzero$ mesons measured in the same $\pt$ interval~\cite{RaaD} to obtain the extrapolated $\Ds$ yield, which amounts to about 70\% of the total production yield. An additional uncertainty was assigned to the extrapolation procedure, by repeating the computation using the TAMU and Catania transport models, and the SHMc to extrapolate the $\Ds/\Dzero$ ratio in the unmeasured $\pt$ interval. Finally, the $\pt$-integrated production yield was obtained as the sum of the extrapolated one for $\pt<2~\GeV/c$ and the measured one. The results for the 0--10\% and 30--50\% centrality intervals are reported in Table~\ref{tab:ptint_yields}. As for the $\Dzero$, $\Dplus$, and $\Dstar$ mesons~\cite{RaaD}, the production yield of prompt $\Ds$ mesons at midrapidity is compatible within uncertainties with the one predicted by the SHMc. This suggests that low-$\pt$ charm quarks, which determine the total yield, are thermalised in the QGP.

\begin{table}[!tb]
\caption{Production yields of prompt $\Ds$ mesons in $|y| < 0.5$ in Pb--Pb collisions at $\sqrtsNN=5.02~\TeV$ compared to the predictions of the GSI-Heidelberg SHMc~\cite{Andronic:2021erx}.}
\centering
\renewcommand*{\arraystretch}{1.4}
\begin{tabular}[t]{l|>{\centering}p{0.5\linewidth}>{\centering\arraybackslash}p{0.3\linewidth}}
\toprule
Centrality & $\de N/\de y|_{|y|<0.5}$ & GSI-Heidelberg SHMc \\
\midrule
0--10\%  & $1.89 \pm 0.07(\mathrm{stat}) ^{+0.13}_{-0.16} (\mathrm{syst}) ^{+0.36}_{-0.55} (\mathrm{extr})\pm 0.07(\mathrm{BR})$ & $2.22 \pm 0.38$\\
30--50\% & $0.34 \pm 0.01(\mathrm{stat}) ^{+0.02}_{-0.03} (\mathrm{syst}) ^{+0.11}_{-0.09} (\mathrm{extr})\pm 0.01(\mathrm{BR})$ & $0.344\pm0.056$ \\
\bottomrule
\end{tabular}
\label{tab:ptint_yields}	
\end{table}

\begin{figure}[!b]
    \begin{center}
    \includegraphics[width = 0.9\textwidth]{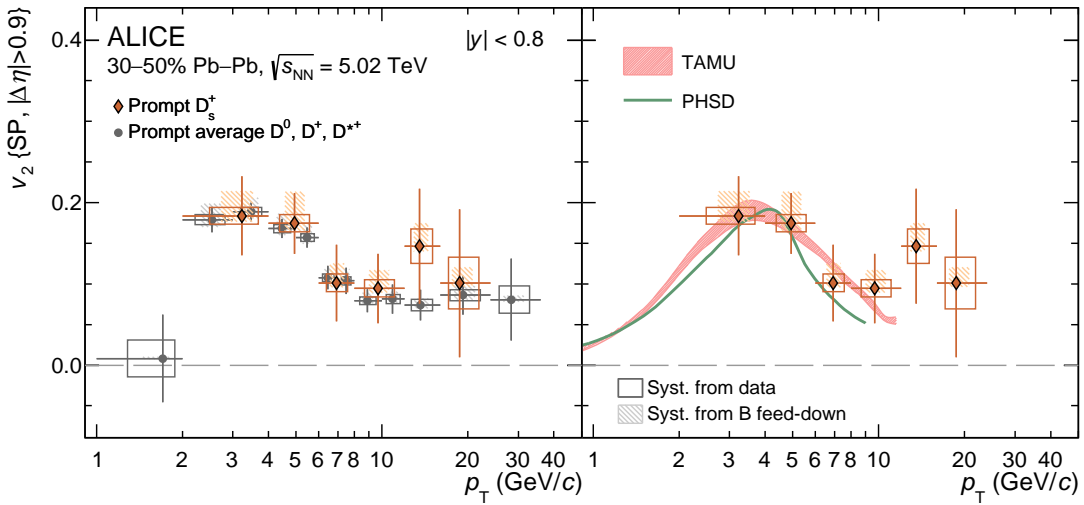}
    \end{center}
    \caption{Elliptic flow coefficient $v_2$ of prompt $\Ds$ mesons in the 30--50\% centrality interval in Pb--Pb collisions at $\sqrtsNN=5.02~\TeV$ compared with that of non-strange D mesons~\cite{Acharya:2020pnh} (left panel) and with theoretical calculations based on the charm-quark transport in a hydrodynamically expanding QGP~\cite{Song:2015sfa, He:2019vgs} (right panel).}
    \label{fig:v2}
\end{figure}

The degree of thermalisation of charm quarks and their hadronisation in the QGP were also studied via the measurement of the azimuthal anisotropy in the prompt $\Ds$-meson production. Figure~\ref{fig:v2} shows the elliptic flow coefficient $v_2$ of prompt $\Ds$ mesons for the 30--50\% centrality interval measured in the transverse-momentum interval $2<\pt<24~\GeV/c$, compared with that of prompt non-strange D mesons (left panel) and with theoretical calculations (right panel). The rapidity interval of the measurement, $|y|<0.8$, is wider than that quoted for the $\RAA$ since no correction for the rapidity acceptance was applied. The measurement was carried out in finer $\pt$ intervals and has uncertainties reduced by a factor up to four with respect to the previous measurement~\cite{Acharya:2017qps}, thanks to the more advanced $\Ds$-meson selection technique and the larger data sample. Considering as null hypothesis $v_2=0$, the probability to observe the measured positive $v_2$ in $2<\pt<8~\GeV/c$ corresponds to a significance of $6.4\sigma$, confirming the participation of the charm quark in the collective motion of the system, as already observed for non-strange D mesons~\cite{Acharya:2017qps,Acharya:2020pnh}. However, within the current uncertainties it is not possible to draw a conclusion about a potential difference between the elliptic flow of strange and non-strange D mesons, which would be motivated by the different mass, the charm-quark hadronisation via recombination with strange quarks in the medium instead of light quarks~\cite{Molnar:2004ph}, and possible differences in the hadronic phase~\cite{He:2012df}. The measured $\Ds$-meson $v_2$ is compatible within uncertainties with the predictions of the TAMU and PHSD models, which include  charm-quark coalescence with flowing strange quarks in the medium.

%% file: conclusions.tex
\section{Summary}

In this Letter, a comprehensive and high-precision set of measurements regarding the prompt $\Ds$-meson production at midrapidity in Pb--Pb collisions at $\sqrtsNN=5.02~\TeV$ was reported. 

The $\pt$-differential production yields were measured in a wide transverse-momentum interval between 2--50 (2--36) $\GeV/c$ in the 0--10\% (30--50\%) centrality interval. They were used to compute the $\pt$-differential $\RAA$ and the ratio of $\Ds$-meson production relative to $\Dzero$ mesons. 
The measured $\RAA$ shows a strong suppression of the $\Ds$-meson production yield compared to the binary-scaled pp reference, reaching a minimum of about 0.2 (0.4) around $\pt\approx10~\GeV/c$ in the 0--10\% (30--50\%) centrality interval. For lower $\pt$, the $\RAA$ increases reaching about unity for $\pt\approx 2\text{--}3~\GeV/c$. The $\Ds/\Dzero$ yield ratios in Pb--Pb collisions are higher than those measured in pp collisions for $\pt \lesssim 8~\GeV/c$ with a significance of $2.3\sigma$ and $2.4\sigma$ in the 0--10\% and 30--50\% centrality intervals, respectively. This finding is consistent with the predictions of theoretical calculations implementing the charm-quark transport in a hydrodynamically expanding QGP, which include an enhanced strange-quark production in the medium and the charm-quark hadronisation via coalescence. The production yield of prompt $\Ds$ mesons, extrapolated down to $\pt=0$, in the 0--10\% centrality interval is compatible with the prediction of the SHMc, suggesting that the bulk of charm quarks are thermalised in the QGP.

The elliptic flow coefficient $v_2$ of prompt $\Ds$ mesons was measured as a function of $\pt$ in the 30--50\% centrality interval. The $\Ds$-meson $v_2$ in $2<\pt<8~\GeV/c$ is positive with a significance of $6.4\sigma$ and is compatible within uncertainties with that of non-strange D mesons. The measured $v_2$ is also described by several transport-model calculations implementing the charm-quark hadronisation via coalescence.

The data reported in this Letter represent the most precise measurements of prompt $\Ds$-meson production in heavy-ion collisions at LHC energies to date, and provide stringent constraints to all models on the production of charm quarks and their hadronisation in the QGP. Future data samples that will be collected with the upgraded ALICE detector in Run 3 will have the potential to further improve and extend to lower $\pt$ the measurement of $\Ds$ mesons in heavy-ion collisions~\cite{Citron:2018lsq}.

%% file: fa_2021-08-20.tex

The ALICE Collaboration would like to thank all its engineers and technicians for their invaluable contributions to the construction of the experiment and the CERN accelerator teams for the outstanding performance of the LHC complex.
The ALICE Collaboration gratefully acknowledges the resources and support provided by all Grid centres and the Worldwide LHC Computing Grid (WLCG) collaboration.
The ALICE Collaboration acknowledges the following funding agencies for their support in building and running the ALICE detector:
A. I. Alikhanyan National Science Laboratory (Yerevan Physics Institute) Foundation (ANSL), State Committee of Science and World Federation of Scientists (WFS), Armenia;
Austrian Academy of Sciences, Austrian Science Fund (FWF): [M 2467-N36] and Nationalstiftung f\"{u}r Forschung, Technologie und Entwicklung, Austria;
Ministry of Communications and High Technologies, National Nuclear Research Center, Azerbaijan;
Conselho Nacional de Desenvolvimento Cient\'{\i}fico e Tecnol\'{o}gico (CNPq), Financiadora de Estudos e Projetos (Finep), Funda\c{c}\~{a}o de Amparo \`{a} Pesquisa do Estado de S\~{a}o Paulo (FAPESP) and Universidade Federal do Rio Grande do Sul (UFRGS), Brazil;
Ministry of Education of China (MOEC) , Ministry of Science \& Technology of China (MSTC) and National Natural Science Foundation of China (NSFC), China;
Ministry of Science and Education and Croatian Science Foundation, Croatia;
Centro de Aplicaciones Tecnol\'{o}gicas y Desarrollo Nuclear (CEADEN), Cubaenerg\'{\i}a, Cuba;
Ministry of Education, Youth and Sports of the Czech Republic, Czech Republic;
The Danish Council for Independent Research | Natural Sciences, the VILLUM FONDEN and Danish National Research Foundation (DNRF), Denmark;
Helsinki Institute of Physics (HIP), Finland;
Commissariat \`{a} l'Energie Atomique (CEA) and Institut National de Physique Nucl\'{e}aire et de Physique des Particules (IN2P3) and Centre National de la Recherche Scientifique (CNRS), France;
Bundesministerium f\"{u}r Bildung und Forschung (BMBF) and GSI Helmholtzzentrum f\"{u}r Schwerionenforschung GmbH, Germany;
General Secretariat for Research and Technology, Ministry of Education, Research and Religions, Greece;
National Research, Development and Innovation Office, Hungary;
Department of Atomic Energy Government of India (DAE), Department of Science and Technology, Government of India (DST), University Grants Commission, Government of India (UGC) and Council of Scientific and Industrial Research (CSIR), India;
Indonesian Institute of Science, Indonesia;
Istituto Nazionale di Fisica Nucleare (INFN), Italy;
Japanese Ministry of Education, Culture, Sports, Science and Technology (MEXT), Japan Society for the Promotion of Science (JSPS) KAKENHI and Japanese Ministry of Education, Culture, Sports, Science and Technology (MEXT)of Applied Science (IIST), Japan;
Consejo Nacional de Ciencia (CONACYT) y Tecnolog\'{i}a, through Fondo de Cooperaci\'{o}n Internacional en Ciencia y Tecnolog\'{i}a (FONCICYT) and Direcci\'{o}n General de Asuntos del Personal Academico (DGAPA), Mexico;
Nederlandse Organisatie voor Wetenschappelijk Onderzoek (NWO), Netherlands;
The Research Council of Norway, Norway;
Commission on Science and Technology for Sustainable Development in the South (COMSATS), Pakistan;
Pontificia Universidad Cat\'{o}lica del Per\'{u}, Peru;
Ministry of Education and Science, National Science Centre and WUT ID-UB, Poland;
Korea Institute of Science and Technology Information and National Research Foundation of Korea (NRF), Republic of Korea;
Ministry of Education and Scientific Research, Institute of Atomic Physics and Ministry of Research and Innovation and Institute of Atomic Physics, Romania;
Joint Institute for Nuclear Research (JINR), Ministry of Education and Science of the Russian Federation, National Research Centre Kurchatov Institute, Russian Science Foundation and Russian Foundation for Basic Research, Russia;
Ministry of Education, Science, Research and Sport of the Slovak Republic, Slovakia;
National Research Foundation of South Africa, South Africa;
Swedish Research Council (VR) and Knut \& Alice Wallenberg Foundation (KAW), Sweden;
European Organization for Nuclear Research, Switzerland;
Suranaree University of Technology (SUT), National Science and Technology Development Agency (NSDTA) and Office of the Higher Education Commission under NRU project of Thailand, Thailand;
Turkish Energy, Nuclear and Mineral Research Agency (TENMAK), Turkey;
National Academy of  Sciences of Ukraine, Ukraine;
Science and Technology Facilities Council (STFC), United Kingdom;
National Science Foundation of the United States of America (NSF) and United States Department of Energy, Office of Nuclear Physics (DOE NP), United States of America.

%% file: 2021-08-20-Alice_Authorlist_2021-08-20.tex
\small
\begin{flushleft}

\bigskip 

S.~Acharya$^{\rm 143}$, 
D.~Adamov\'{a}$^{\rm 98}$, 
A.~Adler$^{\rm 76}$, 
J.~Adolfsson$^{\rm 83}$, 
G.~Aglieri Rinella$^{\rm 35}$, 
M.~Agnello$^{\rm 31}$, 
N.~Agrawal$^{\rm 55}$, 
Z.~Ahammed$^{\rm 143}$, 
S.~Ahmad$^{\rm 16}$, 
S.U.~Ahn$^{\rm 78}$
% ...

\bigskip

\bigskip 

\textbf{\Large Affiliation Notes}

\bigskip 

% ...
$^{\rm VII}$ Also at: University of Kansas, Lawrence, Kansas, United States\\

\bigskip

\bigskip 

\textbf{\Large Collaboration Institutes}

\bigskip 

% ...
$^{16}$ Department of Physics, Aligarh Muslim University, Aligarh, India\\
% ...
$^{31}$ Dipartimento DISAT del Politecnico and Sezione INFN, Turin, Italy\\
% ...
$^{35}$ European Organization for Nuclear Research (CERN), Geneva, Switzerland\\
% ...
$^{55}$ INFN, Sezione di Bologna, Bologna, Italy\\
% ...
$^{76}$ Johann-Wolfgang-Goethe Universit\"{a}t Frankfurt Institut f\"{u}r Informatik, Fachbereich Informatik und Mathematik, Frankfurt, Germany\\
% ...
$^{78}$ Korea Institute of Science and Technology Information, Daejeon, Republic of Korea\\
% ...
$^{83}$ Lund University Department of Physics, Division of Particle Physics, Lund, Sweden\\
% ...
$^{98}$ Nuclear Physics Institute of the Czech Academy of Sciences, \v{R}e\v{z} u Prahy, Czech Republic\\
% ...
$^{143}$ Variable Energy Cyclotron Centre, Homi Bhabha National Institute, Kolkata, India\\
% ...

\end{flushleft} 

%% file: promptDs.bbl
\providecommand{\href}[2]{#2}\begingroup\raggedright\begin{thebibliography}{10}

\bibitem{Bazavov:2018mes}
{\bfseries HotQCD} Collaboration, A.~Bazavov {\em et~al.}, ``{Chiral crossover
  in QCD at zero and non-zero chemical potentials}'',
  \href{http://dx.doi.org/10.1016/j.physletb.2019.05.013}{{\em Phys. Lett. B}
  {\bfseries 795} (2019) 15--21},
  \href{http://arxiv.org/abs/1812.08235}{{\ttfamily arXiv:1812.08235
  [hep-lat]}}.

\bibitem{Borsanyi:2020fev}
S.~Borsanyi, Z.~Fodor, J.~N. Guenther, R.~Kara, S.~D. Katz, P.~Parotto,
  A.~Pasztor, C.~Ratti, and K.~K. Szabo, ``{QCD Crossover at Finite Chemical
  Potential from Lattice Simulations}'',
  \href{http://dx.doi.org/10.1103/PhysRevLett.125.052001}{{\em Phys. Rev.
  Lett.} {\bfseries 125} no.~5, (2020) 052001},
  \href{http://arxiv.org/abs/2002.02821}{{\ttfamily arXiv:2002.02821
  [hep-lat]}}.

\bibitem{Busza:2018rrf}
W.~Busza, K.~Rajagopal, and W.~van~der Schee, ``{Heavy Ion Collisions: The Big
  Picture, and the Big Questions}'',
  \href{http://dx.doi.org/10.1146/annurev-nucl-101917-020852}{{\em Ann. Rev.
  Nucl. Part. Sci.} {\bfseries 68} (2018) 339--376},
  \href{http://arxiv.org/abs/1802.04801}{{\ttfamily arXiv:1802.04801
  [hep-ph]}}.

\bibitem{Aamodt:2011mr}
{\bfseries ALICE} Collaboration, K.~Aamodt {\em et~al.}, ``{Two-pion
  Bose-Einstein correlations in central Pb-Pb collisions at $\sqrt{{s}_{NN}} =$
  2.76 TeV}'', \href{http://dx.doi.org/10.1016/j.physletb.2010.12.053}{{\em
  Phys. Lett. B} {\bfseries 696} (2011) 328--337},
  \href{http://arxiv.org/abs/1012.4035}{{\ttfamily arXiv:1012.4035 [nucl-ex]}}.

\bibitem{Moore:2004tg}
G.~D. Moore and D.~Teaney, ``{How much do heavy quarks thermalize in a heavy
  ion collision?}'', \href{http://dx.doi.org/10.1103/PhysRevC.71.064904}{{\em
  Phys. Rev.} {\bfseries C71} (2005) 064904},
\href{http://arxiv.org/abs/hep-ph/0412346}{{\ttfamily arXiv:hep-ph/0412346
  [hep-ph]}}.

\bibitem{Rapp:2008qc}
R.~Rapp and H.~van Hees, ``{Heavy Quark Diffusion as a Probe of the Quark-Gluon
  Plasma}'', \href{http://arxiv.org/abs/0803.0901}{{\ttfamily arXiv:0803.0901
  [hep-ph]}}.

\bibitem{Batsouli:2002qf}
S.~Batsouli, S.~Kelly, M.~Gyulassy, and J.~L. Nagle, ``{Does the charm flow at
  RHIC?}'', \href{http://dx.doi.org/10.1016/S0370-2693(03)00175-8}{{\em Phys.
  Lett. B} {\bfseries 557} (2003) 26--32},
  \href{http://arxiv.org/abs/nucl-th/0212068}{{\ttfamily
  arXiv:nucl-th/0212068}}.

\bibitem{Fries:2003vb}
R.~Fries, B.~Muller, C.~Nonaka, and S.~Bass, ``{Hadronization in heavy ion
  collisions: Recombination and fragmentation of partons}'',
  \href{http://dx.doi.org/10.1103/PhysRevLett.90.202303}{{\em Phys. Rev. Lett.}
  {\bfseries 90} (2003) 202303},
  \href{http://arxiv.org/abs/nucl-th/0301087}{{\ttfamily
  arXiv:nucl-th/0301087}}.

\bibitem{Greco:2003xt}
V.~Greco, C.~Ko, and P.~Levai, ``{Parton coalescence and anti-proton / pion
  anomaly at RHIC}'',
  \href{http://dx.doi.org/10.1103/PhysRevLett.90.202302}{{\em Phys. Rev. Lett.}
  {\bfseries 90} (2003) 202302},
  \href{http://arxiv.org/abs/nucl-th/0301093}{{\ttfamily
  arXiv:nucl-th/0301093}}.

\bibitem{Greco:2003vf}
V.~Greco, C.~M. Ko, and R.~Rapp, ``{Quark coalescence for charmed mesons in
  ultrarelativistic heavy ion collisions}'',
  \href{http://dx.doi.org/10.1016/j.physletb.2004.06.064}{{\em Phys. Lett. B}
  {\bfseries 595} (2004) 202--208},
  \href{http://arxiv.org/abs/nucl-th/0312100}{{\ttfamily
  arXiv:nucl-th/0312100}}.

\bibitem{Ravagli:2007xx}
L.~Ravagli and R.~Rapp, ``{Quark Coalescence based on a Transport Equation}'',
  \href{http://dx.doi.org/10.1016/j.physletb.2007.07.043}{{\em Phys. Lett. B}
  {\bfseries 655} (2007) 126--131},
  \href{http://arxiv.org/abs/0705.0021}{{\ttfamily arXiv:0705.0021 [hep-ph]}}.

\bibitem{Lisovyi:2015uqa}
M.~Lisovyi, A.~Verbytskyi, and O.~Zenaiev, ``{Combined analysis of charm-quark
  fragmentation-fraction measurements}'',
  \href{http://dx.doi.org/10.1140/epjc/s10052-016-4246-y}{{\em Eur. Phys. J. C}
  {\bfseries 76} no.~7, (2016) 397},
  \href{http://arxiv.org/abs/1509.01061}{{\ttfamily arXiv:1509.01061
  [hep-ex]}}.

\bibitem{Acharya:2021set}
{\bfseries ALICE} Collaboration, S.~Acharya {\em et~al.}, ``{Charm-quark
  fragmentation fractions and production cross section at midrapidity in pp
  collisions at the LHC}'', \href{http://arxiv.org/abs/2105.06335}{{\ttfamily
  arXiv:2105.06335 [nucl-ex]}}.

\bibitem{Wang:1998bha}
X.-N. Wang, ``{Effect of jet quenching on high $p_{T}$ hadron spectra in
  high-energy nuclear collisions}'',
  \href{http://dx.doi.org/10.1103/PhysRevC.58.2321}{{\em Phys. Rev. C}
  {\bfseries 58} (1998) 2321},
  \href{http://arxiv.org/abs/hep-ph/9804357}{{\ttfamily arXiv:hep-ph/9804357}}.

\bibitem{Miller:2007ri}
M.~L. Miller, K.~Reygers, S.~J. Sanders, and P.~Steinberg, ``{Glauber modeling
  in high energy nuclear collisions}'',
  \href{http://dx.doi.org/10.1146/annurev.nucl.57.090506.123020}{{\em Ann. Rev.
  Nucl. Part. Sci.} {\bfseries 57} (2007) 205--243},
  \href{http://arxiv.org/abs/nucl-ex/0701025}{{\ttfamily
  arXiv:nucl-ex/0701025}}.

\bibitem{Ollitrault:1992bk}
J.-Y. Ollitrault, ``{Anisotropy as a signature of transverse collective
  flow}'', \href{http://dx.doi.org/10.1103/PhysRevD.46.229}{{\em Phys. Rev. D}
  {\bfseries 46} (1992) 229--245}.

\bibitem{Poskanzer:1998yz}
A.~M. Poskanzer and S.~A. Voloshin, ``{Methods for analyzing anisotropic flow
  in relativistic nuclear collisions}'',
  \href{http://dx.doi.org/10.1103/PhysRevC.58.1671}{{\em Phys. Rev. C}
  {\bfseries 58} (1998) 1671--1678},
  \href{http://arxiv.org/abs/nucl-ex/9805001}{{\ttfamily
  arXiv:nucl-ex/9805001}}.

\bibitem{Adam:2018inb}
{\bfseries STAR} Collaboration, J.~Adam {\em et~al.}, ``{Centrality and
  transverse momentum dependence of $D^0$-meson production at mid-rapidity in
  Au+Au collisions at ${\sqrt{s_{\rm NN}} = \rm{200\,GeV}}$}'',
  \href{http://dx.doi.org/10.1103/PhysRevC.99.034908}{{\em Phys. Rev. C}
  {\bfseries 99} no.~3, (2019) 034908},
  \href{http://arxiv.org/abs/1812.10224}{{\ttfamily arXiv:1812.10224
  [nucl-ex]}}.

\bibitem{Sirunyan:2017xss}
{\bfseries CMS} Collaboration, A.~M. Sirunyan {\em et~al.}, ``{Nuclear
  modification factor of D$^0$ mesons in PbPb collisions at
  $\sqrt{s_\mathrm{NN}} = 5.02$ TeV}'',
  \href{http://dx.doi.org/10.1016/j.physletb.2018.05.074}{{\em Phys. Lett. B}
  {\bfseries 782} (2018) 474--496},
  \href{http://arxiv.org/abs/1708.04962}{{\ttfamily arXiv:1708.04962
  [nucl-ex]}}.

\bibitem{Acharya:2018hre}
{\bfseries ALICE} Collaboration, S.~Acharya {\em et~al.}, ``{Measurement of
  D$^{0}$, D$^{+}$, D$^{*+}$ and D$_{s}^{+}$ production in Pb-Pb collisions at
  $ \sqrt{{\mathrm{s}}_{\mathrm{NN}}}=5.02 $ TeV}'',
  \href{http://dx.doi.org/10.1007/JHEP10(2018)174}{{\em JHEP} {\bfseries 10}
  (2018) 174}, \href{http://arxiv.org/abs/1804.09083}{{\ttfamily
  arXiv:1804.09083 [nucl-ex]}}.

\bibitem{RaaD}
{\bfseries ALICE} Collaboration, S.~Acharya {\em et~al.}, ``{Prompt D$^{0}$,
  D$^{+}$, and D$^{*+}$ production in Pb-Pb collisions at $\sqrt{s_{\rm NN}}$ =
  5.02 TeV}'', \href{http://arxiv.org/abs/2110.09420}{{\ttfamily
  arXiv:2110.09420 [nucl-ex]}}.

\bibitem{Xu:2014tda}
J.~Xu, J.~Liao, and M.~Gyulassy, ``{Consistency of Perfect Fluidity and Jet
  Quenching in semi-Quark-Gluon Monopole Plasmas}'',
  \href{http://dx.doi.org/10.1088/0256-307X/32/9/092501}{{\em Chin. Phys.
  Lett.} {\bfseries 32} no.~9, (2015) 092501},
  \href{http://arxiv.org/abs/1411.3673}{{\ttfamily arXiv:1411.3673 [hep-ph]}}.

\bibitem{Stojku:2020tuk}
S.~Stojku, B.~Ilic, M.~Djordjevic, and M.~Djordjevic, ``{Extracting the
  temperature dependence in high-$p_\perp$ particle energy loss}'',
  \href{http://dx.doi.org/10.1103/PhysRevC.103.024908}{{\em Phys. Rev. C}
  {\bfseries 103} no.~2, (2021) 024908},
  \href{http://arxiv.org/abs/2007.07851}{{\ttfamily arXiv:2007.07851
  [nucl-th]}}.

\bibitem{Kang:2016ofv}
Z.-B. Kang, F.~Ringer, and I.~Vitev, ``{Effective field theory approach to open
  heavy flavor production in heavy-ion collisions}'',
  \href{http://dx.doi.org/10.1007/JHEP03(2017)146}{{\em JHEP} {\bfseries 03}
  (2017) 146}, \href{http://arxiv.org/abs/1610.02043}{{\ttfamily
  arXiv:1610.02043 [hep-ph]}}.

\bibitem{Sirunyan:2017plt}
{\bfseries CMS} Collaboration, A.~M. Sirunyan {\em et~al.}, ``{Measurement of
  prompt $D^0$ meson azimuthal anisotropy in Pb-Pb collisions at
  $\sqrt{{s}_{NN}}$ = 5.02 TeV}'',
  \href{http://dx.doi.org/10.1103/PhysRevLett.120.202301}{{\em Phys. Rev.
  Lett.} {\bfseries 120} no.~20, (2018) 202301},
  \href{http://arxiv.org/abs/1708.03497}{{\ttfamily arXiv:1708.03497
  [nucl-ex]}}.

\bibitem{Acharya:2020pnh}
{\bfseries ALICE} Collaboration, S.~Acharya {\em et~al.},
  ``{Transverse-momentum and event-shape dependence of D-meson flow harmonics
  in Pb\textendash{}Pb collisions at $\sqrt {s_{NN}}$ = 5.02 TeV}'',
  \href{http://dx.doi.org/10.1016/j.physletb.2020.136054}{{\em Phys. Lett. B}
  {\bfseries 813} (2021) 136054},
  \href{http://arxiv.org/abs/2005.11131}{{\ttfamily arXiv:2005.11131
  [nucl-ex]}}.

\bibitem{Nahrgang:2013xaa}
M.~Nahrgang, J.~Aichelin, P.~B. Gossiaux, and K.~Werner, ``{Influence of
  hadronic bound states above $T_c$ on heavy-quark observables in Pb + Pb
  collisions at at the CERN Large Hadron Collider}'',
  \href{http://dx.doi.org/10.1103/PhysRevC.89.014905}{{\em Phys. Rev. C}
  {\bfseries 89} no.~1, (2014) 014905},
  \href{http://arxiv.org/abs/1305.6544}{{\ttfamily arXiv:1305.6544 [hep-ph]}}.

\bibitem{Katz:2019fkc}
R.~Katz, C.~A.~G. Prado, J.~Noronha-Hostler, J.~Noronha, and A.~A.~P. Suaide,
  ``{Sensitivity study with a D and B mesons modular simulation code of heavy
  flavor RAA and azimuthal anisotropies based on beam energy, initial
  conditions, hadronization, and suppression mechanisms}'',
  \href{http://dx.doi.org/10.1103/PhysRevC.102.024906}{{\em Phys. Rev. C}
  {\bfseries 102} no.~2, (2020) 024906},
  \href{http://arxiv.org/abs/1906.10768}{{\ttfamily arXiv:1906.10768
  [nucl-th]}}.

\bibitem{Beraudo:2014boa}
A.~Beraudo, A.~De~Pace, M.~Monteno, M.~Nardi, and F.~Prino, ``{Heavy flavors in
  heavy-ion collisions: quenching, flow and correlations}'',
  \href{http://dx.doi.org/10.1140/epjc/s10052-015-3336-6}{{\em Eur. Phys. J. C}
  {\bfseries 75} no.~3, (2015) 121},
  \href{http://arxiv.org/abs/1410.6082}{{\ttfamily arXiv:1410.6082 [hep-ph]}}.

\bibitem{Beraudo:2017gxw}
A.~Beraudo, A.~De~Pace, M.~Monteno, M.~Nardi, and F.~Prino, ``{Development of
  heavy-flavour flow-harmonics in high-energy nuclear collisions}'',
  \href{http://dx.doi.org/10.1007/JHEP02(2018)043}{{\em JHEP} {\bfseries 02}
  (2018) 043}, \href{http://arxiv.org/abs/1712.00588}{{\ttfamily
  arXiv:1712.00588 [hep-ph]}}.

\bibitem{Cao:2016gvr}
S.~Cao, T.~Luo, G.-Y. Qin, and X.-N. Wang, ``{Linearized Boltzmann transport
  model for jet propagation in the quark-gluon plasma: Heavy quark
  evolution}'', \href{http://dx.doi.org/10.1103/PhysRevC.94.014909}{{\em Phys.
  Rev. C} {\bfseries 94} no.~1, (2016) 014909},
  \href{http://arxiv.org/abs/1605.06447}{{\ttfamily arXiv:1605.06447
  [nucl-th]}}.

\bibitem{Cao:2017hhk}
S.~Cao, T.~Luo, G.-Y. Qin, and X.-N. Wang, ``{Heavy and light flavor jet
  quenching at RHIC and LHC energies}'',
  \href{http://dx.doi.org/10.1016/j.physletb.2017.12.023}{{\em Phys. Lett. B}
  {\bfseries 777} (2018) 255--259},
  \href{http://arxiv.org/abs/1703.00822}{{\ttfamily arXiv:1703.00822
  [nucl-th]}}.

\bibitem{He:2019vgs}
M.~He and R.~Rapp, ``{Hadronization and Charm-Hadron Ratios in Heavy-Ion
  Collisions}'', \href{http://dx.doi.org/10.1103/PhysRevLett.124.042301}{{\em
  Phys. Rev. Lett.} {\bfseries 124} no.~4, (2020) 042301},
  \href{http://arxiv.org/abs/1905.09216}{{\ttfamily arXiv:1905.09216
  [nucl-th]}}.

\bibitem{Li:2019lex}
S.~Li and J.~Liao, ``{Data-driven extraction of heavy quark diffusion in
  quark-gluon plasma}'',
  \href{http://dx.doi.org/10.1140/epjc/s10052-020-8243-9}{{\em Eur. Phys. J. C}
  {\bfseries 80} no.~7, (2020) 671},
  \href{http://arxiv.org/abs/1912.08965}{{\ttfamily arXiv:1912.08965
  [hep-ph]}}.

\bibitem{Scardina:2017ipo}
F.~Scardina, S.~K. Das, V.~Minissale, S.~Plumari, and V.~Greco, ``{Estimating
  the charm quark diffusion coefficient and thermalization time from D meson
  spectra at energies available at the BNL Relativistic Heavy Ion Collider and
  the CERN Large Hadron Collider}'',
  \href{http://dx.doi.org/10.1103/PhysRevC.96.044905}{{\em Phys. Rev. C}
  {\bfseries 96} no.~4, (2017) 044905},
  \href{http://arxiv.org/abs/1707.05452}{{\ttfamily arXiv:1707.05452
  [nucl-th]}}.

\bibitem{Plumari:2019hzp}
S.~Plumari, G.~Coci, V.~Minissale, S.~K. Das, Y.~Sun, and V.~Greco, ``{Heavy -
  light flavor correlations of anisotropic flows at LHC energies within
  event-by-event transport approach}'',
  \href{http://dx.doi.org/10.1016/j.physletb.2020.135460}{{\em Phys. Lett. B}
  {\bfseries 805} (2020) 135460},
  \href{http://arxiv.org/abs/1912.09350}{{\ttfamily arXiv:1912.09350
  [hep-ph]}}.

\bibitem{Ke:2018jem}
W.~Ke, Y.~Xu, and S.~A. Bass, ``{Modified Boltzmann approach for modeling the
  splitting vertices induced by the hot QCD medium in the deep
  Landau-Pomeranchuk-Migdal region}'',
  \href{http://dx.doi.org/10.1103/PhysRevC.100.064911}{{\em Phys. Rev. C}
  {\bfseries 100} no.~6, (2019) 064911},
  \href{http://arxiv.org/abs/1810.08177}{{\ttfamily arXiv:1810.08177
  [nucl-th]}}.

\bibitem{Song:2015sfa}
T.~Song, H.~Berrehrah, D.~Cabrera, J.~M. Torres-Rincon, L.~Tolos, W.~Cassing,
  and E.~Bratkovskaya, ``{Tomography of the Quark-Gluon-Plasma by Charm
  Quarks}'', \href{http://dx.doi.org/10.1103/PhysRevC.92.014910}{{\em Phys.
  Rev. C} {\bfseries 92} no.~1, (2015) 014910},
  \href{http://arxiv.org/abs/1503.03039}{{\ttfamily arXiv:1503.03039
  [nucl-th]}}.

\bibitem{Rafelski:1982pu}
J.~Rafelski and B.~Muller, ``{Strangeness Production in the Quark - Gluon
  Plasma}'', \href{http://dx.doi.org/10.1103/PhysRevLett.48.1066}{{\em Phys.
  Rev. Lett.} {\bfseries 48} (1982) 1066}. [Erratum: Phys.Rev.Lett. 56, 2334
  (1986)].

\bibitem{Koch:1986ud}
P.~Koch, B.~Muller, and J.~Rafelski, ``{Strangeness in Relativistic Heavy Ion
  Collisions}'', \href{http://dx.doi.org/10.1016/0370-1573(86)90096-7}{{\em
  Phys. Rept.} {\bfseries 142} (1986) 167--262}.

\bibitem{Andronic:2003zv}
A.~Andronic, P.~Braun-Munzinger, K.~Redlich, and J.~Stachel, ``{Statistical
  hadronization of charm in heavy ion collisions at SPS, RHIC and LHC}'',
  \href{http://dx.doi.org/10.1016/j.physletb.2003.07.066}{{\em Phys. Lett. B}
  {\bfseries 571} (2003) 36--44},
  \href{http://arxiv.org/abs/nucl-th/0303036}{{\ttfamily
  arXiv:nucl-th/0303036}}.

\bibitem{Kuznetsova:2006bh}
I.~Kuznetsova and J.~Rafelski, ``{Heavy flavor hadrons in statistical
  hadronization of strangeness-rich QGP}'',
  \href{http://dx.doi.org/10.1140/epjc/s10052-007-0268-9}{{\em Eur. Phys. J. C}
  {\bfseries 51} (2007) 113--133},
  \href{http://arxiv.org/abs/hep-ph/0607203}{{\ttfamily arXiv:hep-ph/0607203}}.

\bibitem{He:2012df}
M.~He, R.~J. Fries, and R.~Rapp, ``{$\mathbf{D_s}$-Meson as Quantitative Probe
  of Diffusion and Hadronization in Nuclear Collisions}'',
  \href{http://dx.doi.org/10.1103/PhysRevLett.110.112301}{{\em Phys. Rev.
  Lett.} {\bfseries 110} no.~11, (2013) 112301},
  \href{http://arxiv.org/abs/1204.4442}{{\ttfamily arXiv:1204.4442 [nucl-th]}}.

\bibitem{Adam:2021qty}
{\bfseries STAR} Collaboration, J.~Adam {\em et~al.}, ``{Observation of
  $D_{s}^{\pm}/D^0$ enhancement in Au+Au collisions at $\sqrt{s_{_{\rm NN}}}$ =
  200 GeV}'', \href{http://arxiv.org/abs/2101.11793}{{\ttfamily
  arXiv:2101.11793 [hep-ex]}}.

\bibitem{Adam:2015jda}
{\bfseries ALICE} Collaboration, J.~Adam {\em et~al.}, ``{Measurement of
  D$_{s}^{+}$ production and nuclear modification factor in Pb-Pb collisions at
  $ \sqrt{{\mathrm{s}}_{\mathrm{NN}}}=$ 2.76 TeV}'',
  \href{http://dx.doi.org/10.1007/JHEP03(2016)082}{{\em JHEP} {\bfseries 03}
  (2016) 082}, \href{http://arxiv.org/abs/1509.07287}{{\ttfamily
  arXiv:1509.07287 [nucl-ex]}}.

\bibitem{Zhao:2018jlw}
J.~Zhao, S.~Shi, N.~Xu, and P.~Zhuang, ``{Sequential Coalescence with Charm
  Conservation in High Energy Nuclear Collisions}'',
  \href{http://arxiv.org/abs/1805.10858}{{\ttfamily arXiv:1805.10858
  [hep-ph]}}.

\bibitem{Plumari:2017ntm}
S.~Plumari, V.~Minissale, S.~K. Das, G.~Coci, and V.~Greco, ``{Charmed Hadrons
  from Coalescence plus Fragmentation in relativistic nucleus-nucleus
  collisions at RHIC and LHC}'',
  \href{http://dx.doi.org/10.1140/epjc/s10052-018-5828-7}{{\em Eur. Phys. J. C}
  {\bfseries 78} no.~4, (2018) 348},
  \href{http://arxiv.org/abs/1712.00730}{{\ttfamily arXiv:1712.00730
  [hep-ph]}}.

\bibitem{Acharya:2017qps}
{\bfseries ALICE} Collaboration, S.~Acharya {\em et~al.}, ``{D-meson azimuthal
  anisotropy in midcentral Pb-Pb collisions at $\mathbf{\sqrt{s_{\rm
  NN}}=5.02}$ TeV}'',
  \href{http://dx.doi.org/10.1103/PhysRevLett.120.102301}{{\em Phys. Rev.
  Lett.} {\bfseries 120} no.~10, (2018) 102301},
  \href{http://arxiv.org/abs/1707.01005}{{\ttfamily arXiv:1707.01005
  [nucl-ex]}}.

\bibitem{Aamodt:2008zz}
{\bfseries ALICE} Collaboration, K.~Aamodt {\em et~al.}, ``{The ALICE
  experiment at the CERN LHC}'',
\href{http://dx.doi.org/10.1088/1748-0221/3/08/S08002}{{\em JINST} {\bfseries
  3} (2008) S08002}.

\bibitem{Abelev:2014ffa}
{\bfseries ALICE} Collaboration, B.~Abelev {\em et~al.}, ``{Performance of the
  ALICE Experiment at the CERN LHC}'',
  \href{http://dx.doi.org/10.1142/S0217751X14300440}{{\em Int. J. Mod. Phys.}
  {\bfseries A29} (2014) 1430044},
\href{http://arxiv.org/abs/1402.4476}{{\ttfamily arXiv:1402.4476 [nucl-ex]}}.

\bibitem{Adam:2015ptt}
{\bfseries ALICE} Collaboration, J.~Adam {\em et~al.}, ``{Centrality dependence
  of the charged-particle multiplicity density at midrapidity in Pb-Pb
  collisions at $\sqrt{s_{\rm NN}}$ = 5.02 TeV}'',
  \href{http://dx.doi.org/10.1103/PhysRevLett.116.222302}{{\em Phys. Rev.
  Lett.} {\bfseries 116} no.~22, (2016) 222302},
  \href{http://arxiv.org/abs/1512.06104}{{\ttfamily arXiv:1512.06104
  [nucl-ex]}}.

\bibitem{ALICE-PUBLIC-2018-011}
{\bfseries ALICE} Collaboration, ``{Centrality determination in heavy ion
  collisions}'', {\em ALICE-PUBLIC-2018-011} (Aug, 2018) .
  \url{https://cds.cern.ch/record/2636623}.

\bibitem{PhysRevD.44.3501}
X.-N. Wang and M.~Gyulassy, ``Hijing: A monte carlo model for multiple jet
  production in $\mathrm{pp}$, $\mathrm{pA}$, and $\mathrm{AA}$ collisions'',
  \href{http://dx.doi.org/10.1103/PhysRevD.44.3501}{{\em Phys. Rev. D}
  {\bfseries 44} (Dec, 1991) 3501--3516}.

\bibitem{Sjostrand:2006za}
T.~Sjostrand, S.~Mrenna, and P.~Z. Skands, ``{PYTHIA 6.4 Physics and Manual}'',
  \href{http://dx.doi.org/10.1088/1126-6708/2006/05/026}{{\em JHEP} {\bfseries
  05} (2006) 026}, \href{http://arxiv.org/abs/hep-ph/0603175}{{\ttfamily
  arXiv:hep-ph/0603175}}.

\bibitem{Sjostrand:2014zea}
T.~Sj\"ostrand, S.~Ask, J.~R. Christiansen, R.~Corke, N.~Desai, P.~Ilten,
  S.~Mrenna, S.~Prestel, C.~O. Rasmussen, and P.~Z. Skands, ``{An introduction
  to PYTHIA 8.2}'', \href{http://dx.doi.org/10.1016/j.cpc.2015.01.024}{{\em
  Comput. Phys. Commun.} {\bfseries 191} (2015) 159--177},
  \href{http://arxiv.org/abs/1410.3012}{{\ttfamily arXiv:1410.3012 [hep-ph]}}.

\bibitem{Skands:2014pea}
P.~Skands, S.~Carrazza, and J.~Rojo, ``{Tuning PYTHIA 8.1: the Monash 2013
  Tune}'', \href{http://dx.doi.org/10.1140/epjc/s10052-014-3024-y}{{\em Eur.
  Phys. J. C} {\bfseries 74} no.~8, (2014) 3024},
  \href{http://arxiv.org/abs/1404.5630}{{\ttfamily arXiv:1404.5630 [hep-ph]}}.

\bibitem{Brun:1994aa}
R.~Brun, F.~Bruyant, F.~Carminati, S.~Giani, M.~Maire, A.~McPherson,
  G.~Patrick, and L.~Urban,
  \href{http://dx.doi.org/10.17181/CERN.MUHF.DMJ1}{{\em {GEANT: Detector
  Description and Simulation Tool; Oct 1994}}}.
\newblock CERN Program Library. CERN, Geneva, 1993.
\newblock \url{http://cds.cern.ch/record/1082634}.
\newblock Long Writeup W5013.

\bibitem{Zyla:2020zbs}
{\bfseries Particle Data Group} Collaboration, P.~Zyla {\em et~al.}, ``{Review
  of Particle Physics}'', \href{http://dx.doi.org/10.1093/ptep/ptaa104}{{\em
  PTEP} {\bfseries 2020} no.~8, (2020) 083C01}.

\bibitem{hipe4ml}
L.~Barioglio, F.~Catalano, M.~Concas, P.~Fecchio, F.~Grosa, F.~Mazzaschi, and
  M.~Puccio, ``hipe4ml/hipe4ml'', July, 2021.
\newblock \url{https://doi.org/10.5281/zenodo.5070132}.

\bibitem{Chen:2016XST}
T.~Chen and C.~Guestrin, ``Xgboost: A scalable tree boosting system'',
  \href{http://dx.doi.org/10.1145/2939672.2939785}{{\em Proceedings of the 22nd
  ACM SIGKDD International Conference on Knowledge Discovery and Data Mining}
  (2016) 785–794}, \href{http://arxiv.org/abs/1603.02754}{{\ttfamily
  arXiv:1603.02754 [cs.LG]}}.

\bibitem{Cacciari:1998it}
M.~Cacciari, M.~Greco, and P.~Nason, ``{The $\pt$ spectrum in heavy-flavor
  hadroproduction}'',
  \href{http://dx.doi.org/10.1088/1126-6708/1998/05/007}{{\em JHEP} {\bfseries
  05} (1998) 007},
\href{http://arxiv.org/abs/hep-ph/9803400}{{\ttfamily arXiv:hep-ph/9803400
  [hep-ph]}}.

\bibitem{Cacciari:2001td}
M.~Cacciari, S.~Frixione, and P.~Nason, ``{The $\pt$ spectrum in heavy-flavor
  photoproduction}'',
  \href{http://dx.doi.org/10.1088/1126-6708/2001/03/006}{{\em JHEP} {\bfseries
  03} (2001) 006},
\href{http://arxiv.org/abs/hep-ph/0102134}{{\ttfamily arXiv:hep-ph/0102134
  [hep-ph]}}.

\bibitem{Adam:2015sza}
{\bfseries ALICE} Collaboration, J.~Adam {\em et~al.}, ``{Transverse momentum
  dependence of D-meson production in Pb-Pb collisions at $
  \sqrt{{\mathrm{s}}_{\mathrm{NN}}}=$ 2.76 TeV}'',
  \href{http://dx.doi.org/10.1007/JHEP03(2016)081}{{\em JHEP} {\bfseries 03}
  (2016) 081}, \href{http://arxiv.org/abs/1509.06888}{{\ttfamily
  arXiv:1509.06888 [nucl-ex]}}.

\bibitem{Acharya:2021cqv}
{\bfseries ALICE} Collaboration, S.~Acharya {\em et~al.}, ``{Measurement of
  beauty and charm production in pp collisions at $ \sqrt{s} $ = 5.02 TeV via
  non-prompt and prompt D mesons}'',
  \href{http://dx.doi.org/10.1007/JHEP05(2021)220}{{\em JHEP} {\bfseries 05}
  (2021) 220}, \href{http://arxiv.org/abs/2102.13601}{{\ttfamily
  arXiv:2102.13601 [nucl-ex]}}.

\bibitem{Voloshin:2008dg}
S.~A. Voloshin, A.~M. Poskanzer, and R.~Snellings, ``{Collective phenomena in
  non-central nuclear collisions}'',
  \href{http://dx.doi.org/10.1007/978-3-642-01539-7_10}{{\em Landolt-Bornstein}
  {\bfseries 23} (2010) 293--333},
  \href{http://arxiv.org/abs/0809.2949}{{\ttfamily arXiv:0809.2949 [nucl-ex]}}.

\bibitem{Luzum:2012da}
M.~Luzum and J.-Y. Ollitrault, ``{Eliminating experimental bias in
  anisotropic-flow measurements of high-energy nuclear collisions}'',
  \href{http://dx.doi.org/10.1103/PhysRevC.87.044907}{{\em Phys. Rev. C}
  {\bfseries 87} no.~4, (2013) 044907},
  \href{http://arxiv.org/abs/1209.2323}{{\ttfamily arXiv:1209.2323 [nucl-ex]}}.

\bibitem{Selyuzhenkov:2007zi}
I.~Selyuzhenkov and S.~Voloshin, ``{Effects of non-uniform acceptance in
  anisotropic flow measurement}'',
  \href{http://dx.doi.org/10.1103/PhysRevC.77.034904}{{\em Phys. Rev. C}
  {\bfseries 77} (2008) 034904},
  \href{http://arxiv.org/abs/0707.4672}{{\ttfamily arXiv:0707.4672 [nucl-th]}}.

\bibitem{Borghini:2004ra}
N.~Borghini and J.~Y. Ollitrault, ``{Azimuthally sensitive correlations in
  nucleus-nucleus collisions}'',
  \href{http://dx.doi.org/10.1103/PhysRevC.70.064905}{{\em Phys. Rev. C}
  {\bfseries 70} (2004) 064905},
  \href{http://arxiv.org/abs/nucl-th/0407041}{{\ttfamily
  arXiv:nucl-th/0407041}}.

\bibitem{ATLAS:2018xms}
{\bfseries ATLAS} Collaboration, M.~Aaboud {\em et~al.}, ``{Prompt and
  non-prompt $J/\psi $ elliptic flow in Pb+Pb collisions at $\sqrt{s_{_\text
  {NN}}} = 5.02$ Tev with the ATLAS detector}'',
  \href{http://dx.doi.org/10.1140/epjc/s10052-018-6243-9}{{\em Eur. Phys. J. C}
  {\bfseries 78} no.~9, (2018) 784},
  \href{http://arxiv.org/abs/1807.05198}{{\ttfamily arXiv:1807.05198
  [nucl-ex]}}.

\bibitem{Khachatryan:2016ypw}
{\bfseries CMS} Collaboration, V.~Khachatryan {\em et~al.}, ``{Suppression and
  azimuthal anisotropy of prompt and nonprompt ${\mathrm{J}}/\psi $ production
  in PbPb collisions at $\sqrt{{s_{_{\text {NN}}}}} =2.76$ $\,\mathrm{TeV}$}'',
  \href{http://dx.doi.org/10.1140/epjc/s10052-017-4781-1}{{\em Eur. Phys. J. C}
  {\bfseries 77} no.~4, (2017) 252},
  \href{http://arxiv.org/abs/1610.00613}{{\ttfamily arXiv:1610.00613
  [nucl-ex]}}.

\bibitem{Uphoff:2012gb}
J.~Uphoff, O.~Fochler, Z.~Xu, and C.~Greiner, ``{Open Heavy Flavor in Pb+Pb
  Collisions at $\sqrt{s}=2.76$ TeV within a Transport Model}'',
  \href{http://dx.doi.org/10.1016/j.physletb.2012.09.069}{{\em Phys. Lett. B}
  {\bfseries 717} (2012) 430--435},
  \href{http://arxiv.org/abs/1205.4945}{{\ttfamily arXiv:1205.4945 [hep-ph]}}.

\bibitem{Aichelin:2012ww}
J.~Aichelin, P.~B. Gossiaux, and T.~Gousset, ``{Radiative and Collisional
  Energy Loss of Heavy Quarks in Deconfined Matter}'',
  \href{http://dx.doi.org/10.5506/APhysPolB.43.655}{{\em Acta Phys. Polon. B}
  {\bfseries 43} (2012) 655--662},
  \href{http://arxiv.org/abs/1201.4192}{{\ttfamily arXiv:1201.4192 [nucl-th]}}.

\bibitem{Greco:2007sz}
V.~Greco, H.~van Hees, and R.~Rapp, ``{Heavy-quark kinetics at RHIC and LHC}'',
  in {\em {23rd International Nuclear Physics Conference (INPC 2007)}}.
\newblock 9, 2007.
\newblock \href{http://arxiv.org/abs/0709.4452}{{\ttfamily arXiv:0709.4452
  [hep-ph]}}.

\bibitem{ALICE-PUBLIC-2017-005}
{\bfseries ALICE} Collaboration, ``{The ALICE definition of primary
  particles}'', {\em ALICE-PUBLIC-2017-005} (Jun, 2017) .
  \url{https://cds.cern.ch/record/2270008}.

\bibitem{Song:2015ykw}
T.~Song, H.~Berrehrah, D.~Cabrera, W.~Cassing, and E.~Bratkovskaya, ``{Charm
  production in Pb + Pb collisions at energies available at the CERN Large
  Hadron Collider}'', \href{http://dx.doi.org/10.1103/PhysRevC.93.034906}{{\em
  Phys. Rev. C} {\bfseries 93} no.~3, (2016) 034906},
  \href{http://arxiv.org/abs/1512.00891}{{\ttfamily arXiv:1512.00891
  [nucl-th]}}.

\bibitem{Cacciari:2012ny}
M.~Cacciari, S.~Frixione, N.~Houdeau, M.~L. Mangano, P.~Nason, and G.~Ridolfi,
  ``{Theoretical predictions for charm and bottom production at the LHC}'',
  \href{http://dx.doi.org/10.1007/JHEP10(2012)137}{{\em JHEP} {\bfseries 10}
  (2012) 137}, \href{http://arxiv.org/abs/1205.6344}{{\ttfamily arXiv:1205.6344
  [hep-ph]}}.

\bibitem{ALICE-PUBLIC-2018-014}
{\bfseries ALICE} Collaboration, ``{ALICE 2017 luminosity determination for pp
  collisions at $\sqrt{s}$ = 5 TeV}'', {\em ALICE-PUBLIC-2018-014} (Nov, 2018)
  . \url{http://cds.cern.ch/record/2648933}.

\bibitem{Berrehrah:2016vzw}
H.~Berrehrah, E.~Bratkovskaya, T.~Steinert, and W.~Cassing, ``{A dynamical
  quasiparticle approach for the QGP bulk and transport properties}'',
  \href{http://dx.doi.org/10.1142/S0218301316420039}{{\em Int. J. Mod. Phys. E}
  {\bfseries 25} no.~07, (2016) 1642003},
  \href{http://arxiv.org/abs/1605.02371}{{\ttfamily arXiv:1605.02371
  [hep-ph]}}.

\bibitem{Dover:1991zn}
C.~B. Dover, U.~W. Heinz, E.~Schnedermann, and J.~Zimanyi, ``{Relativistic
  coalescence model for high-energy nuclear collisions}'',
  \href{http://dx.doi.org/10.1103/PhysRevC.44.1636}{{\em Phys. Rev. C}
  {\bfseries 44} (1991) 1636--1654}.

\bibitem{Minissale:2020bif}
V.~Minissale, S.~Plumari, and V.~Greco, ``{Charm Hadrons in pp collisions at
  LHC energy within a Coalescence plus Fragmentation approach}'',
  \href{http://arxiv.org/abs/2012.12001}{{\ttfamily arXiv:2012.12001
  [hep-ph]}}.

\bibitem{He:2019tik}
M.~He and R.~Rapp, ``{Charm-Baryon Production in Proton-Proton Collisions}'',
  \href{http://dx.doi.org/10.1016/j.physletb.2019.06.004}{{\em Phys. Lett. B}
  {\bfseries 795} (2019) 117--121},
  \href{http://arxiv.org/abs/1902.08889}{{\ttfamily arXiv:1902.08889
  [nucl-th]}}.

\bibitem{Andronic:2021erx}
A.~Andronic, P.~Braun-Munzinger, M.~K. K\"ohler, A.~Mazeliauskas, K.~Redlich,
  J.~Stachel, and V.~Vislavicius, ``{The multiple-charm hierarchy in the
  statistical hadronization model}'',
  \href{http://dx.doi.org/10.1007/JHEP07(2021)035}{{\em JHEP} {\bfseries 07}
  (2021) 035}, \href{http://arxiv.org/abs/2104.12754}{{\ttfamily
  arXiv:2104.12754 [hep-ph]}}.

\bibitem{Mazeliauskas:2018irt}
A.~Mazeliauskas, S.~Floerchinger, E.~Grossi, and D.~Teaney, ``{Fast resonance
  decays in nuclear collisions}'',
  \href{http://dx.doi.org/10.1140/epjc/s10052-019-6791-7}{{\em Eur. Phys. J. C}
  {\bfseries 79} no.~3, (2019) 284},
  \href{http://arxiv.org/abs/1809.11049}{{\ttfamily arXiv:1809.11049
  [nucl-th]}}.

\bibitem{Molnar:2004ph}
D.~Molnar, ``{Charm elliptic flow from quark coalescence dynamics}'',
  \href{http://dx.doi.org/10.1088/0954-3899/31/4/052}{{\em J. Phys. G}
  {\bfseries 31} (2005) S421--S428},
  \href{http://arxiv.org/abs/nucl-th/0410041}{{\ttfamily
  arXiv:nucl-th/0410041}}.

\bibitem{Citron:2018lsq}
Z.~Citron {\em et~al.}, ``{Report from Working Group 5}: {Future physics
  opportunities for high-density QCD at the LHC with heavy-ion and proton
  beams}'', \href{http://dx.doi.org/10.23731/CYRM-2019-007.1159}{{\em CERN
  Yellow Rep. Monogr.} {\bfseries 7} (2019) 1159--1410},
  \href{http://arxiv.org/abs/1812.06772}{{\ttfamily arXiv:1812.06772
  [hep-ph]}}.

\end{thebibliography}\endgroup
